\documentclass[a4paper]{iopart}  

\usepackage[T1]{fontenc} 
\usepackage{textcomp} 
\usepackage{lmodern} 
\usepackage[british]{babel} 
\usepackage[a4paper]{geometry} 
\usepackage[parfill]{parskip}    
\usepackage{subfigure}
\usepackage{graphicx}
\usepackage{amssymb}
\usepackage{epstopdf}

\usepackage{color,soul}
\usepackage[hidelinks=true,bookmarksnumbered=true]{hyperref}

\usepackage{float} 

\newcommand{\Qgb}{Q_{\rm gB}}

\begin{document}

\title{Collisionality scaling of the electron heat flux in ETG turbulence}
\author{G~J~Colyer$^{1,\,2\,3}$, A~A~Schekochihin$^{1,\,4}$, F~I~Parra$^{1,\,2}$, C~M~Roach$^{2}$, 
M~A~Barnes$^{1,\,2,\,5}$, Y-c~Ghim$^{1,\,2,\,6}$ and W~Dorland$^{7,\,1}$}
\address{$^1$ Rudolf Peierls Centre for Theoretical Physics, University of Oxford, OX1 3NP, UK}
\address{$^2$ CCFE, Culham Science Centre, Abingdon, OX14 3DB, UK}
\address{$^3$ 
Engineering, Mathematics and Physical Sciences, University of Exeter, EX4 4QF, UK}
\address{$^4$ Merton College, Oxford, OX1 4JD, UK}
\address{$^5$ Plasma Science and Fusion Center, 167 Albany Street, Cambridge, MA 02139, USA}
\address{$^6$ Department of Nuclear and Quantum Engineering, KAIST, Daejeon, 34141, Korea}
\address{$^7$ Department of Physics, University of Maryland, College Park, MD 20742-4111, USA}

\begin{abstract}
In electrostatic simulations of MAST plasma at electron-gyroradius scales, using the local flux-tube gyrokinetic code GS2 with adiabatic ions, we find that the long-time saturated electron heat flux (the level most relevant to energy transport) decreases as the electron collisionality decreases. At early simulation times, the heat flux ``quasi-saturates'' without any strong dependence on collisionality, and with the turbulence dominated by streamer-like radially elongated structures. However, the zonal fluctuation component continues to grow slowly until much later times, eventually leading to a new saturated state dominated by zonal modes and with the heat flux proportional to the collision rate, in approximate agreement with the experimentally observed collisionality scaling of the energy confinement in MAST. We outline an explanation of this effect based on a model of ETG turbulence dominated by zonal-nonzonal interactions and on an analytically derived scaling of the zonal-mode damping rate with the electron-ion collisionality. Improved energy confinement with decreasing collisionality is favourable towards the performance of future, hotter devices.
\end{abstract}

\section{Introduction}
\label{etg-introduction}

Experiments on MAST in which heat transport is dominated by the electron channel find that the thermal energy confinement time $\tau_E$ varies with the normalised electron collisionality $\nu_{\ast}$ according to the scaling \cite{Valovic}
\begin{equation}
B\tau_E \propto\nu_{\ast}^{-0.82\pm 0.1},
\label{MASTscaling}
\end{equation}
where $B$ is the magnetic field.
This scaling is favourable towards improved confinement in the hotter, lower collisionality plasmas
anticipated in future devices.

In this paper, we investigate how the electron heat flux $Q$ varies with electron collisionality in
simulations of electron-scale plasma turbulence in MAST, using the local gyrokinetic flux-tube code GS2 \cite{KOTSCHENREUTHER1995, GS2}.
At constant geometry and $\rho_\ast = \rho_e/a$ ($\rho_e$ is the electron Larmor radius, $a$ is the equilibrium length scale),
it can be shown (see \ref{scaling_correspondence}) that 
\begin{equation} 
B\tau_E \propto \left(\frac{Q}{\Qgb}\right)^{-1}, 
\end{equation}
where $\Qgb = n T v_{te} \rho_\ast^2$ is the electron gyroBohm heat flux.  
It therefore ought to be possible to recover the scaling (\ref{MASTscaling}) from a local calculation of the electron heat flux.

We wish to discover whether this experimental MAST scaling may be understood in terms of electron
temperature gradient (ETG) driven turbulent transport. 
With this goal in mind, we carry out local gyrokinetic simulations restricted to electrostatic
perturbations at electron-gyroradius scales, ignoring both electromagnetic (e.g., ``microtearing'' 
\cite{Doerk11,Gutten11})
modes and ion-gyroradius-scale effects. The restriction to electrostatic perturbations is a matter of considering as simple a model as possible for a plasma where beta is low and magnetic perturbations are relatively small. The neglect of the ion-gyroradius-scale turbulence is justifiable on the grounds that, 
in typical MAST plasmas, ion turbulence 
is considerably suppressed by radial shear in the background flow, and ion transport is close to the neoclassical
level \cite{Field04, Roach05, FieldIAEA2010, FieldNF2011}. This is fortunate, as spanning both ion- and electron-gyroradius scales requires prohibitively large computational resources.
We further limit
ourselves here to a Boltzmann-ion model, at electron scales, treating only the electrons kinetically. Whilst greatly
simplified, the equations we solve (which are described in more detail in section \ref{setup}) reproduce the experimental scaling of electron heat flux with collisionality.

\begin{figure}[b]
\centering
\subfigure[]{
  \includegraphics[scale=0.65,clip]{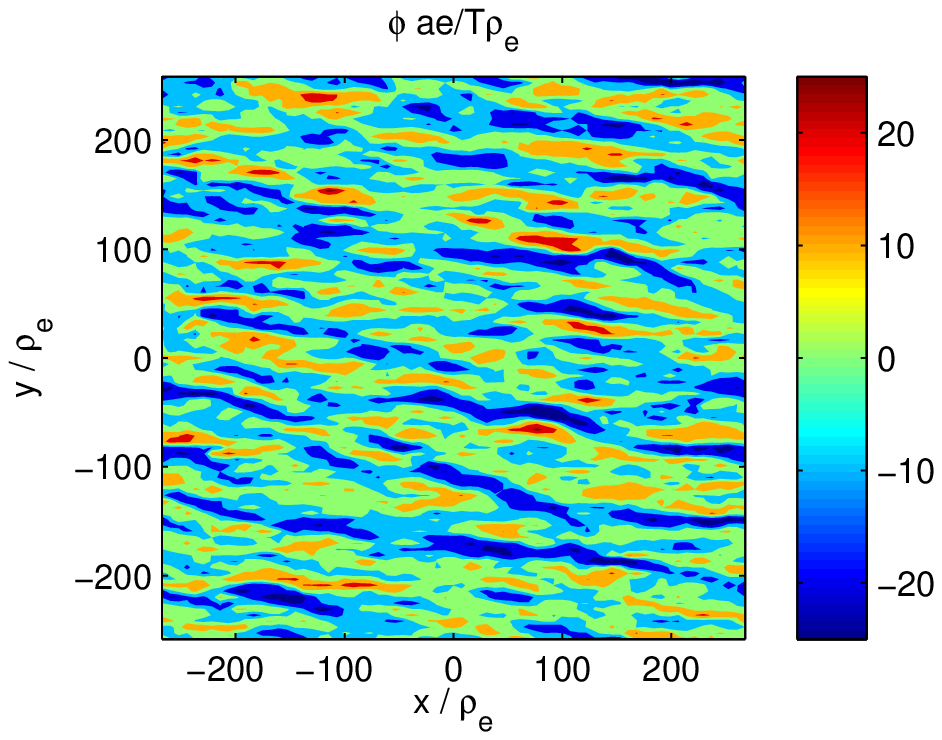} 
\label{quasisatslice}
}
\subfigure[]{
  \includegraphics[scale=0.65,clip]{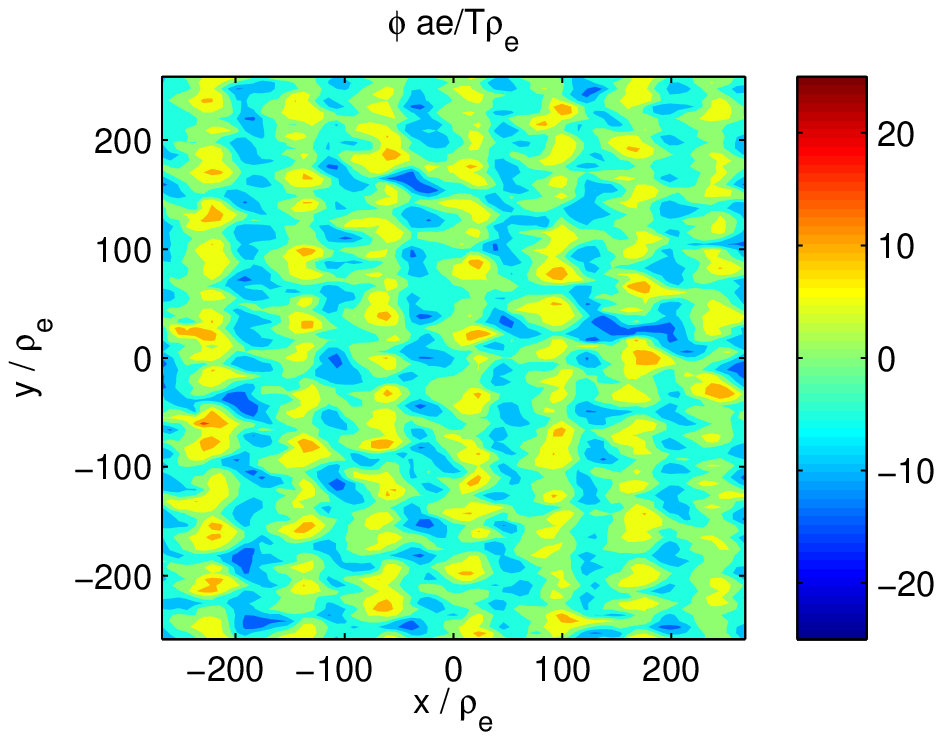} 
\label{satslice}
}
\caption{Non-dimensionalised electrostatic potential $e\phi/T\rho_\ast$ (where $\rho_\ast=\rho_e/a$) at the outboard midplane, for $\nu = 0.2 \,\nu_{\mathrm{nom}}$ (here $\nu_\mathrm{nom}$ is the ``nominal'', i.e., experimental value of collisionality), $a/L_T = 3.3$: \subref{quasisatslice} quasi-saturated state at $t = 1200.3\,a/v_{te}$, \subref{satslice} saturated state at $t = 7835.8\,a/v_{te}$, for large-box simulations. See \ref{plasma_parameters} and \ref{numerical_parameters} for the meaning of symbols.}
\label{contours}
\end{figure}

The reference simulation parameters are based on experiment, which is close to the threshold for the onset of turbulence. In this region of parameter space, we find that the saturated turbulent heat flux varies with collisionality in a manner consistent with the experimental scaling. This scaling is only revealed, however, if the simulation times are sufficiently long to reach a true steady state, which requires them to be much longer than the electron-collision time scale. At earlier times, there is a transient ``quasi-saturated'' state with higher heat flux, in which the zonal modes (which do not themselves contribute to the heat flux) are small but slowly growing. When they have grown to a sufficient level, the nonzonal modes and the heat flux are significantly suppressed.
Figure \ref{contours} illustrates these two regimes by showing the electrostatic potential (which is proportional to the density perturbation) in an outboard-midplane cross-section of a flux tube in MAST both in the earlier quasi-saturated state and the later long-time saturated state, based on one of the simulations reported below. In the quasi-saturated state, the zonal modes do not appear to play a special role, and radially extended ``streamers'' can be seen, as is usually expected for ETG turbulence \cite{Dorland00, Jenko00}. In contrast, in the long-time saturated state, a strong zonal component comes to dominate, structuring the turbulence into
``vortex streets'' and dramatically weakening radial transport. 
In this final saturated state, the nonlinear drive of the zonal modes is balanced by their weak collisional damping, dominated by electron-ion collisions. Scans in collisionality reveal that the saturated heat flux increases with increasing collisionality, in rough proportionality: $Q/\Qgb\propto\nu_{\ast }$.

A brief outline of the rest of the paper is as follows. In section \ref{setup}, we describe the equations that are solved and the simulation set-up. 
In section \ref{results}, we present our main results, including the long-time evolution of the turbulence, the dependence of the saturated heat flux on collisionality, and the structure of the
saturated turbulent state. 
We also sketch a simple theoretical argument that explains the collisionality scaling
of the heat flux (section \ref{twiddle}). In section \ref{discussion}, a summary of our findings 
is given, our results are put in the context of some earlier work, and a discussion is given 
of the apparent differences and possible similarities between the ETG and ITG turbulent states 
in light of the conclusions of the present study. 

\section{Governing equations and numerical set-up} 
\label{setup}

Our study is based on the MAST H-mode shot 8500, which had 2 MW of NBI heating, and for which data are available
from the International Tokamak Profile Database \cite{Roach08}. This shot was analysed by Field et al. \cite{Field04}, and a linear gyrokinetic study was performed by Roach et al. \cite{Roach05}. In the present work, we consider a single flux surface, for which the detailed plasma parameters are given in \ref{plasma_parameters} (they are referred to as ``nominal'' parameters); these were kept fixed throughout our study, except for varying collisionality and electron temperature gradient where indicated.
In \ref{numerical_parameters}, we provide the technical details about our numerical simulations: the coordinate system used, numerical grids, resolution and boundary conditions.

\subsection{Gyrokinetic equation}

We use the GS2 continuum gyrokinetic code \cite{GS2} to obtain the perturbed distribution function and electrostatic field in local flux-tube geometry.
The electron distribution function is written 
\begin{equation}
f = F + \delta f
\end{equation}
(we omit species subscripts when this will cause no confusion), where
\begin{equation}
F = n\left(\frac{m}{2\pi T}\right)^{3/2}\exp\left(-\frac{mv^2}{2T}\right)
\end{equation}
is the equilibrium Maxwellian background distribution, $n$, $m$, $T$ and $\mathbf v$ are the electron density, mass, temperature and velocity, respectively, and the perturbed distribution function is split into a Boltzmann response associated with the perturbed electrostatic potential $\phi$ and a gyrotropic non-Boltzmann, generally non-Maxwellian part: 
\begin{equation}
\delta f = \frac{e\phi(\mathbf{r})}{T} F+h(\mathbf{R},v_\perp,v_\parallel),
\label{dfe}
\end{equation}
where $e$ is the absolute value of the electron charge. 
Note that $\phi$ is a function of the particle position $\mathbf r$, whereas $h$ is 
a function of the gyrocentre position $\mathbf R=\mathbf r - \mathbf b\times\mathbf v/\Omega_e$, 
where $\mathbf{b}$ is the unit vector along the magnetic field and $\Omega_e=-eB/mc$ is the electron 
cyclotron frequency (its sign is the negative sign of the electron charge). The gyrocentre distribution $h$ is otherwise independent of the gyroangle, 
being a function of the parallel $v_\parallel = {\mathbf v}\cdot\mathbf b$ and 
perpendicular $v_\perp = (v^2 - v_\parallel^2)^{1/2}$ velocities. 
Equivalently, its velocity-space variables can be (and are) chosen to be 
the energy ${\cal E} = v^2/2$ and magnetic moment $\mu = v_\perp^2/2B$, 
so $v_\parallel = \pm(2{\cal E} - 2\mu B)^{1/2}$. In this representation, 
the gyrocentre distribution $h$ satisfies the electrostatic gyrokinetic equation (GKE)
\cite{Frieman82} (see review \cite{Abel13})
\begin{equation}
\frac{\partial}{\partial t}\left(h + \frac{e\langle\phi\rangle}{T}F\right) 
+\left(v_\parallel\mathbf{b}+\mathbf{v}_B\right)\cdot\nabla h 
+\langle\mathbf{v}_E\rangle\cdot\nabla h +\langle\mathbf{v}_E\rangle\cdot\nabla F=
\langle C[h]\rangle,
\label{GKE}
\end{equation} 
where $\langle\dots\rangle$ denotes gyroaveraging at constant gyrocentre position $\mathbf R$, 
\begin{equation}
\mathbf{v}_B= \frac{\mathbf{b}}{\Omega_e}\times\left(v_\parallel^2 \mathbf{b}\cdot\nabla\mathbf{b} 
+ \frac{v_\perp^2}{2}\frac{\nabla B}{B}\right) 
\label{vB}
\end{equation}
is the magnetic drift velocity, and
\begin{equation}
\mathbf{v}_E = \frac{c}{B}\mathbf{b}\times \nabla\phi
\label{ExB}
\end{equation}
is the $\mathbf{E\times B}$ drift velocity. The energy is injected into the system 
via the last term on the left-hand side of (\ref{GKE}), which contains the radial gradients 
of the equilibrium distribution: 
\begin{equation}
\langle\mathbf{v}_E\rangle\cdot\nabla F = 
\langle v_{Ex}\rangle\frac{\partial F}{\partial x}
= \langle v_{Ex}\rangle
\left[\frac{1}{L_n} + \left(\frac{v^2}{v_{te}^2} - \frac{3}{2}\right)\frac{1}{L_T}\right] F,
\end{equation}
where $v_{te}=(2T/m)^{1/2}$ is the electron thermal speed. We have defined  
$L_n= -d\ln n/dx$ and $L_T = -d\ln T/dx$, the 
gradient scale lengths of the equlibrium electron density and temperature profiles, 
respectively. The natural normalisation of these lengths 
is the tokamak minor radius $a$. In the above definitions, 
$x$ is the spatial coordinate transverse to the flux surface 
and $y$ is a coordinate within the flux surface transverse to the 
magnetic field (at the outboard midplane, it is approximately the poloidal arc length) --- these curvilinear, non-orthogonal coordinates, 
as used by GS2, are explained in \ref{coords}.   
Note that, in these coordinates,  
\begin{equation}
v_{Ex} = \mathbf{v}_{E}\cdot\nabla x = \frac{c q}{B_0^2 r}\frac{d\psi}{d r}
\frac{\partial \phi}{\partial y} 
\label{vEx}
\end{equation} 
(see \ref{coords} for the definition of all symbols).

\subsection{Adiabatic ions}

In a simple (two-species, hydrogenic, i.e., with ion charge $=+e$) plasma, the quasineutrality condition for the perturbations 
is 
\begin{equation}
\delta n_e = \delta n_i, 
\label{QN}
\end{equation}
where 
$\delta n = \int d^3\mathbf{v}\delta f$ is the density perturbation for each species. 
In terms of $h$, this density perturbation is 
\begin{equation}
\frac{\delta n}{n} = -\frac{Z e\phi}{T}+\frac{1}{n}\int d^3\mathbf{v}\langle h\rangle_{\mathbf{r}},
\label{dnn}
\end{equation}
where $\langle\cdot\rangle_{\mathbf{r}}$ denotes the gyroaveraging operator at constant 
particle position $\mathbf r$ (needed here because the velocity integral must be performed 
at fixed particle position $\mathbf r$, while $h$ is a function of $\mathbf R$), 
and $Z=1$ for the ions and $-1$ for the electrons.  
We have assumed a decomposition of the perturbed ion distribution function 
analogous to (\ref{dfe}). 
If we consider only perturbations at scales
perpendicular to the magnetic field that are much smaller than the ion gyroradius, 
$k_\perp \rho_i \gg 1$, the non-Boltzmann part of the ion density response 
in (\ref{dnn}) can be neglected because $\langle h_i\rangle_{\mathbf{r}}$
is suppressed by the averaging over large ion Larmor orbits. 
In formal terms, this approximation is the lowest-order
expansion in the electron-ion mass ratio. Therefore, in the present study, we will 
assume that the ion distribution is entirely described by the 
Boltzmann (sometimes called adiabatic) response: 
\begin{equation}
\frac{\delta n_i}{n_i} = -\frac{e\phi}{T_i}. 
\label{ai}
\end{equation}
Combining this with the quasineutrality (\ref{QN}) and the full gyrokinetic 
electron density response given by (\ref{dnn}) with $Z=-1$, 
we get the following equation for $\phi$: 
\begin{equation}
\frac{e\phi}{T_e}\left(1+\frac{1}{\tau}\right) =-\frac{1}{n}\int d^3\mathbf{v}\langle h\rangle_{\mathbf{r}},
\label{QN2}
\end{equation}
where $h$ satisfies (\ref{GKE}) and $\tau=T_i/T_e$. 

Equations (\ref{GKE}) and (\ref{QN2}) constitute a closed system, which is solved numerically 
by the version of the GS2 code that we use here. 

\subsection{Collisions}
\label{colls}

The electron collision operator used in GS2, appearing on the right-hand side of (\ref{GKE}), consists of the electron-ion pitch-angle scattering operator (to lowest order in the mass-ratio expansion) and a particle-, momentum- and energy-conserving simplified model of the electron-electron collision operator \cite{Roach05,Abel08,Barnes09}: in $\mathbf{k}_\perp$ space, it is
\begin{equation}
\fl
\langle C[h]\rangle_{\mathbf{k}_\perp} = 
\langle C_{ee}[h]\rangle_{\mathbf{k}_\perp} + \nu_{ei}\frac{v_{te}^3}{v^3} 
\left[ 
\frac{\partial}{\partial\xi}\frac{(1-\xi^2)}{2}\frac{\partial h_{\mathbf{k}_\perp}}{\partial\xi}
- \frac{(1 + \xi^2)}{4} \frac{v^2}{v_{te}^2} 
k_\perp^2 \rho_e^2 h_{\mathbf{k}_\perp}
\right],
\label{model_e}
\end{equation}
where $\xi = v_\parallel/v$, the electron-electron operator $\langle C_{ee}[h]\rangle$ 
is given in \cite{Abel08} and the numerical implementation of the whole operator is detailed in \cite{Barnes09}.\footnote{As originally written, GS2 was not configured to simulate kinetic electrons 
with adiabatic (Boltzmann) ions and also to include electron-ion collisions. This is because GS2 always formally requires at least one kinetic ion species, and so single-species KE+AI simulations were actually performed as single-species kinetic-ions simulations with the sign of the charge in the Boltzmann response appropriately reversed. The equations solved are then correct for simulations with kinetic electrons and adiabatic ions without collisions or with same-species collisions only. For the present work, we added the capability to include the electron-ion collision term in such simulations; previously it was included only for electrons in multi-species simulations with both kinetic ions and kinetic electrons.} 
The electron-ion collision rate is $\nu_{ei} = Z_{\mathrm{eff}}\nu_{ee}$, where 
$\nu_{ee} \equiv \nu = \sqrt 2 \pi n e^4 T^{-3/2} m^{-1/2} \ln\Lambda$ is the electron-electron collision rate and $\ln\Lambda$ is the Coulomb logarithm. In a plasma with multiple ion species,
$Z_{\mathrm{eff}} = \sum_i n_i Z_i^2/\sum_i n_i Z_i$
arises from summing the individual electron-ion collision operators 
over all ion species ($Z_ie$ is that species' charge); in a hydrogenic plasma, $Z_\mathrm{eff}=1$. Note, however, that GS2 treats $Z_{\mathrm{eff}}$ as an independent parameter in equation (\ref{model_e}), formally allowing one to vary $\nu_{ei}$
relative to $\nu_{ee}$ without affecting the quasineutrality equation (\ref{QN}). 
This is normally done to include the effect of electron collisions with impurity ion species that are not being modelled kinetically, but are present in real experiments. In this work, it will also be useful to us in section \ref{zeffscansection} as a method of varying this ratio artificially. 

Electron-ion collisions relax the parallel electron flow towards the stationary ion background, resistively dissipating the current associated with an electron flow below the ion gyroscale.\footnote{The electron flow is also a current because, in the limit of $k_\perp\rho_i\gg1$, ion gyromotion averages out over the electron scales, leaving, to lowest order, no ion-flow response in either parallel or perpendicular direction; cf.\ the adiabatic-ion approximation (\ref{ai}), which arises from the averaging out of the gyrokinetic ion-density response.} As a result, the electron-ion collision operator, and hence the electron collision operator overall, does not conserve (electron) momentum.\footnote{As we see in section \ref{results}, 
electron-ion collisions reduce the final saturation amplitude of electron-scale zonal flows, and this gives rise to a favourable scaling of electron heat flux with collisionality.}
The last term in the bracket in (\ref{model_e}) is a spatial diffusion --- it is a finite--Larmor-radius (FLR) effect
arising from the gyroaveraging of the collision operator \cite{Abel08}. 

\section{Results}\label{results}

\subsection{Time evolution: from \texorpdfstring{``quasi-saturation''}{"quasi-saturation"} to long-time steady state}

\begin{figure}[t]
\centering
\subfigure[]{ 
\includegraphics[scale=0.9]{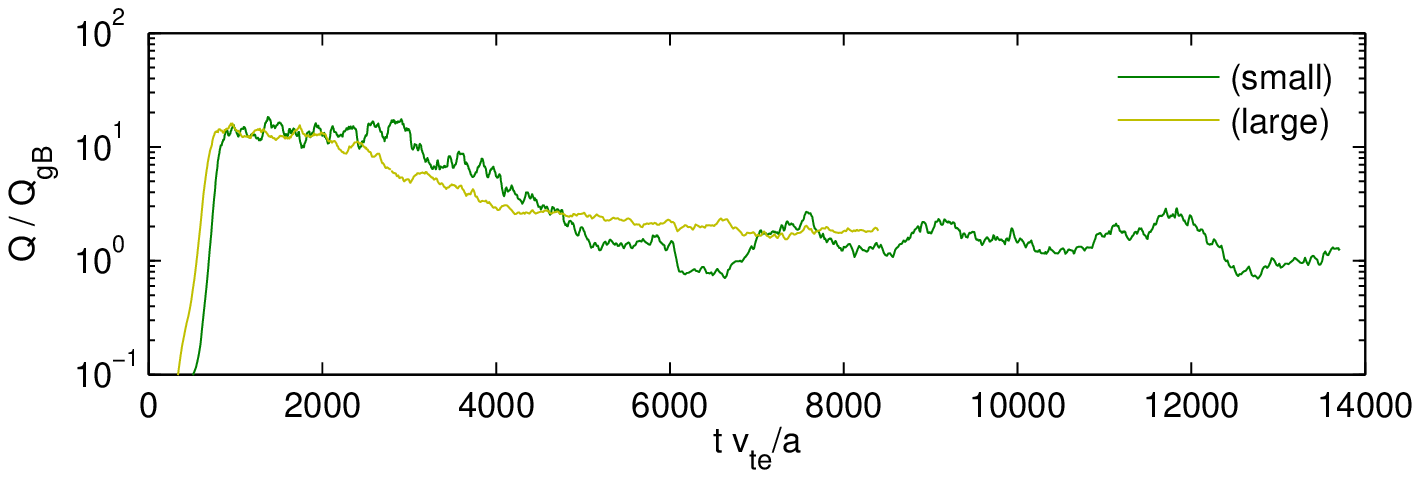} 
\label{qe_overlap}
}
\subfigure[]{
\includegraphics[scale=0.9]{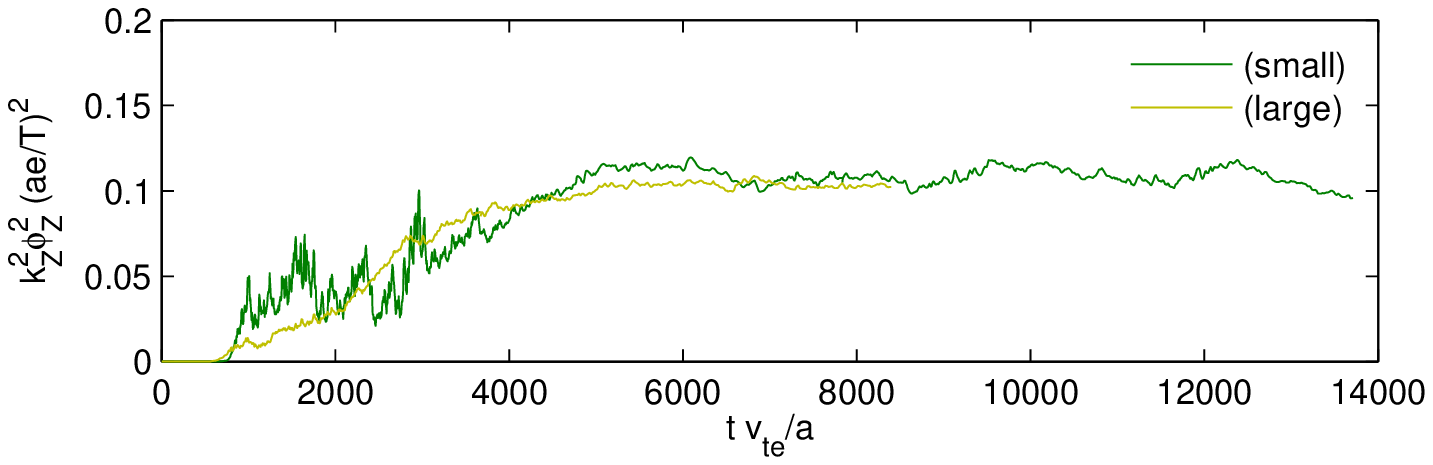}
\label{zv_overlap}
}
\caption{Evolution in time of \subref{qe_overlap} the turbulent electron heat flux in electron gyroBohm units, and \subref{zv_overlap} the square of the zonal velocity, $(k_{\mathrm Z}\phi_{\mathrm Z})^2$, for two simulations with electron collisionality $\nu = 0.2 \,\nu_{\mathrm{nom}}$, electron temperature gradient $a/L_T = 3.3$ (green: small-box simulation; yellow: large-box simulation; see \ref{numerics} for details).} 
\label{overlap}
\end{figure}

Figure \ref{qe_overlap}
shows the turbulent electron heat flux as a function of time for two simulations that have identical
plasma parameters and differ only in numerical grid parameters (their coincidence
is evidence of numerical convergence; for more details of the numerics and of the 
various issues of convergence, see \ref{convergence_study}, which also contains plots for other 
values of $\nu$ and $a/L_T$ and of other zonal quantities). The heat flux $Q$ 
is calculated from the solution of equations (\ref{GKE}) and (\ref{QN2}) according to
\begin{equation}
Q = \overline{\int d^3\mathbf{v}\frac{mv^2}{2} \delta f v_{Ex}}\,,
\label{Qdef}
\end{equation}
where the overbar indicates a flux-surface average. We will normalise $Q$ to the gyroBohm value $\Qgb = n T v_{te} \rho_\ast^2$, where $\rho_\ast = \rho_e/a$.
Besides its physical meaning as the heat flux, 
$Q$ is also a good proxy for the turbulent fluctuation level of the non-zonal 
modes ($k_y\neq 0$; the $k_y=0$ components of $\phi$ or $h$ do not contribute to $Q$). 

After a short exponential transient during the linear growth phase, the system reaches a 
``quasi-saturated'' turbulent state, which, however, is not the
final steady state. The final saturated state is reached much later, after a slow
decline in heat flux accompanied by slow growth of the zonal ($k_y=0$) component of the turbulence: Figure \ref{zv_overlap} shows the evolution of the zonal velocity squared,  
\begin{equation}
(k_{\mathrm Z}\phi_{\mathrm Z})^2 = \sum_{k_x} k_x^2|\phi_{k_x,0}|^2
\label{kzphiz}
\end{equation} 
(this equation also introduces the definition of the characteristic zonal scale $k_\mathrm{Z}$). 
Snapshots of $\phi$, shown in Figure \ref{contours} and corresponding to early and late times in
one of these simulations, are a vivid illustration of the different structure of the ``quasi-saturated'' and the
final steady state: the former resembles the streamer-dominated state usually associated with ETG turbulence
\cite{Dorland00, Jenko00}, whereas the latter is a zonal-mode-dominated state, which we will now proceed 
to analyse. Note that it is this long-time saturated state that matters for determining the level 
of transport because, even though the time for it to emerge is long by the standards of 
electron-gyroscale dynamics, it is still much shorter than the transport time scale in a tokamak.\footnote{Using MAST data from \cite{Valovic}, $\tau_E$ is of order 10 ms, whereas $a/v_{te}$ is of order 10 ns, a factor of $10^6$ smaller. The longest runs in this paper evolve for times only of order $10^4\,a/v_{te}$, so there is still good scale separation. } 
This is because the self-consistent combination of $\delta f$ local (gradient-driven) gyrokinetics 
at the gyroradius scale, and radial transport evolution at the system scale, implies a scale separation 
in both space and time \cite{BarnesTrinity, Abel13}. 

\subsection{Collisionality scaling: numerical results and theory}
\label{twiddle}

\begin{figure}[t]
\centering
\subfigure[]{
  \includegraphics[scale=0.75,clip]{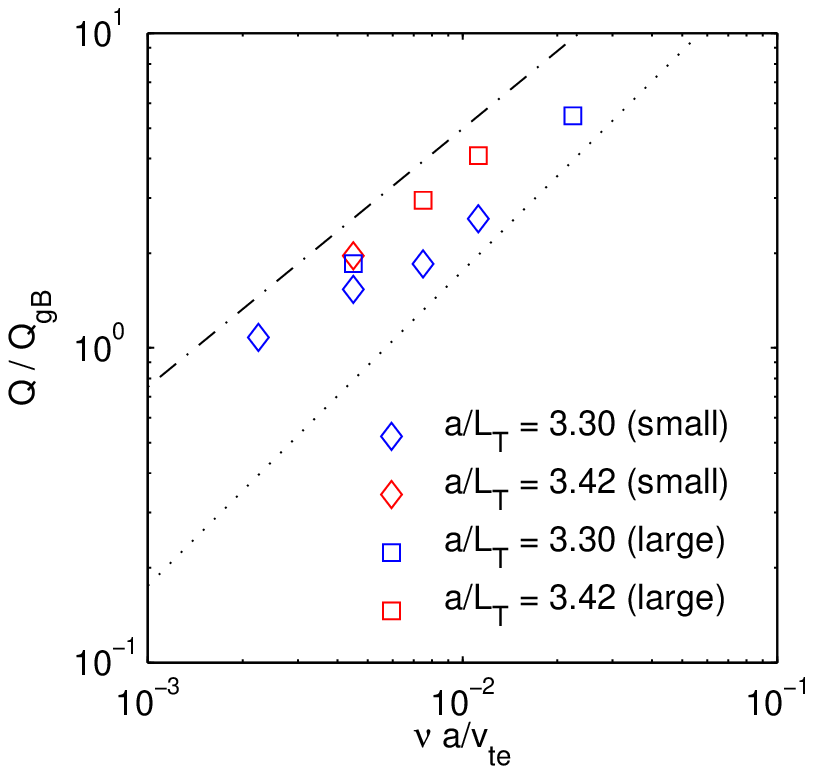} 
  \label{qe_scaling}
}
\subfigure[]{
  \includegraphics[scale=0.75,clip]{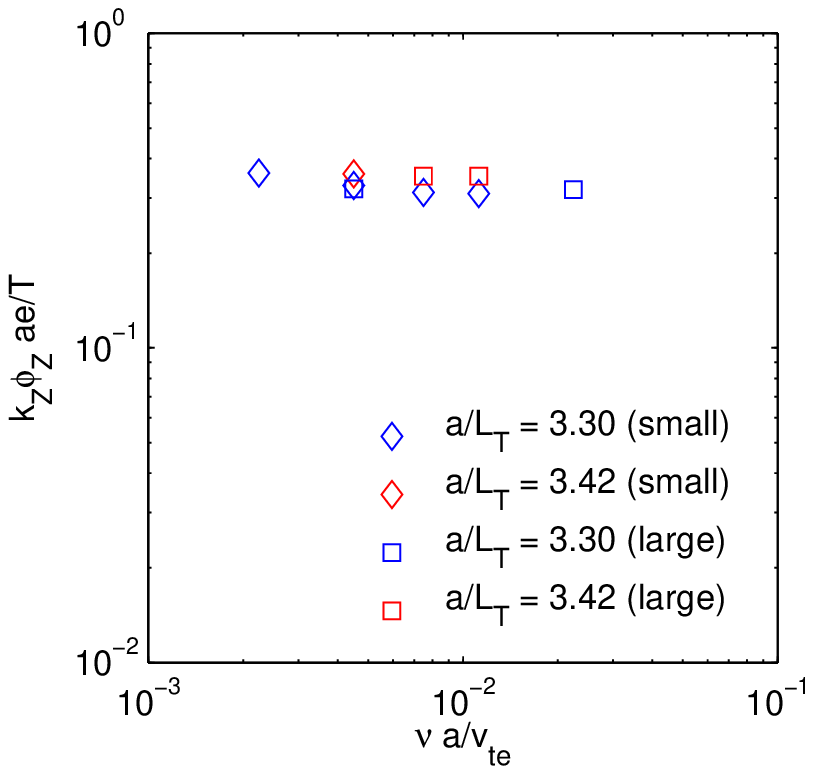} 
  \label{zv_scaling}
}
\caption{Variation of \subref{qe_scaling} the time-averaged normalised electron heat flux $Q/\Qgb$, and \subref{zv_scaling} the rms zonal velocity $k_\mathrm{Z} \phi_\mathrm{Z}$, defined by (\ref{kzphiz}), versus normalised electron collisionality $\nu a/v_{te}$, at the nominal (experimental) value of the temperature gradient $a/L_T = 3.42$, and at $a/L_T = 3.3$. Symbol shapes indicate the simulation series (``small box'' or ``large box''), as explained in \ref{convergence_study}. The dot-dashed line shows the experimental power-law scaling, and
the dotted line shows the theoretical linear scaling, $Q/\Qgb\propto\nu_{\ast}$, equation (\ref{Q_scaling}).}
\label{collscan}
\end{figure}

The two clearest results from the final saturated state of these simulations are summarised in Figure \ref{collscan}, which is based on a parameter scan in electron collisionality for two values of the
electron temperature gradient, at two different numerical resolutions. 
The saturated heat flux increases roughly proportionally to the collisionality, whereas the zonal velocity is essentially independent of it. The heat-flux scaling is in remarkable agreement
with the experimental scaling (\ref{MASTscaling}).\footnote{To facilitate a quantitative comparison of our results with \cite{Valovic} -- whilst noting that we consider a particular flux surface whereas the authors of  \cite{Valovic} considered the plasma globally -- we convert our nominal $\nu = 0.02\,v_{te}/a$ (Table \ref{parameters}) using the expression (\ref{nustar}) to get a nominal $\nu_\ast\approx 0.1$. The collisionality range shown in our Figure \ref{collscan} is thus approximately the same as the range of $\nu_\ast$ accessed experimentally (approximately 0.03--0.1 in Figure 3 of \cite{Valovic}) for which the scaling (\ref{MASTscaling}) was obtained.}

Let us outline a simple explanation of these results, which we will then follow up with 
a series of numerical experiments designed to test its plausibility. 

Let us split the gyrokinetic equation (\ref{GKE}) explicitly into equations
governing the evolution of the
nonzonal and zonal components of the distribution function and the associated 
electrostatic potential (cf.\ \cite{Sugama2009}), 
\begin{equation}
h = h_{\mathrm{NZ}}+h_{\mathrm{Z}}, 
\quad
\phi = \phi_{\mathrm{NZ}}+\phi_{\mathrm{Z}}, 
\end{equation}
where the subscripts NZ, Z denote nonzonal ($k_y\neq0$) and zonal ($k_y=0$) modes, respectively:
\begin{eqnarray}
\fl
\frac{\partial}{\partial t}\left(h_{\mathrm{NZ}} + \frac{e \langle\phi_{\mathrm{NZ}}\rangle}{T} F\right)
+\left(v_\parallel\mathbf{b}+\mathbf{v}_{B}\right)\cdot\nabla h_{\mathrm{NZ}}
- \langle C[h_{\mathrm{NZ}}]\rangle
\nonumber\\
\fl
\qquad + 
\underbrace{\langle\mathbf{v}_{E,{\mathrm{NZ}}}\rangle\cdot\nabla h_{\mathrm{Z}}
+\langle\mathbf{v}_{E,{\mathrm{Z}}}\rangle\cdot\nabla h_{\mathrm{NZ}}}_\mathrm{Z-NZ~interaction~(I)}
+\langle\mathbf{v}_{E,{\mathrm{NZ}}}\rangle\cdot\nabla h_{\mathrm{NZ}}
-\overline{\langle\mathbf{v}_{E,{\mathrm{NZ}}}\rangle\cdot\nabla h_{\mathrm{NZ}}}
\nonumber\\
\fl
\qquad = \underbrace{- \langle\mathbf{v}_{E,{\mathrm{NZ}}}\rangle\cdot\nabla F}_{\mathrm{energy~injection~(II)}},
\label{GKE_NZ}
\\
\fl
\frac{\partial}{\partial t}\left(h_{\mathrm{Z}} + \frac{e \langle\phi_{\mathrm{Z}}\rangle}{T} F\right)
+\left(v_\parallel\mathbf{b}+\mathbf{v}_{B}\right)\cdot\nabla h_{\mathrm{Z}} 
- \underbrace{\langle C[h_{\mathrm{Z}}]\rangle}_\mathrm{damping~(III)}
= \underbrace{- \overline{\langle\mathbf{v}_{E,{\mathrm{NZ}}}\rangle\cdot\nabla h_{\mathrm{NZ}}}}_\mathrm{energy~injection~(IV)},
\label{GKE_Z}
\end{eqnarray}
where the overline denotes spatial averaging over $y$, i.e., the $k_y=0$ component. 
Equation (\ref{GKE_Z}) is the $y$ average of the gyrokinetic equation (\ref{GKE});  
then (\ref{GKE_NZ}) is the result of subtracting (\ref{GKE_Z}) from (\ref{GKE}).  

We conjecture that the dominant balance governing the saturated state of the 
nonzonal modes is between the zonal-nonzonal interaction terms (I) and the 
the linear drive (energy-injection) term (II) in (\ref{GKE_NZ}).\footnote[1]{We are thus treating the collision term in the nonzonal equation as subdominant, or at least as
unimportant to this aspect of the dynamics. Numerically we find that it cannot be neglected as it regularises the fine velocity-space
structure arising due to the phase-mixing of $h_{\mathrm{NZ}}$ \cite{Sch16}. 
The parallel streaming and magnetic-drift terms likely play a part
in determining the spatial structure of the turbulence \cite{Barnes11cb,Ghim13,Sch16}, but we shall see that we do not need to determine $k_y$, $k_{\mathrm{Z}}$ or $k_\parallel$.
In \cite{Sugama2009}, a split between nonzonal and zonal components is performed for integrated entropy balance equations (equations (67) and (68) of \cite{Sugama2009}) in which these other terms do not appear. 
One could base a similar argument to the
one presented here on these equations. 
}
We estimate these terms as 
\begin{equation}
\mathrm{(I)}\sim
\langle\mathbf{v}_{E,{\mathrm{NZ}}}\rangle\cdot\nabla h_{\mathrm{Z}}
\sim \frac{c}{B} k_y\phi_\mathrm{NZ} k_\mathrm{Z} h_\mathrm{Z},
\label{ZNZint}
\end{equation}
and 
\begin{equation}
\mathrm{(II)}\sim
\langle\mathbf{v}_{E,{\mathrm{NZ}}}\rangle\cdot\nabla F 
\sim \frac{c}{B} k_y \phi_\mathrm{NZ} \frac{F}{L_T}
\label{NZdrive}
\end{equation}
where $k_{\mathrm{Z}}$ is the typical zonal wavenumber and $k_y$ the typical nonzonal wavenumber
(we have used $v_{Ex}\sim (c/B) k_y\phi$; see (\ref{vEx})). 
The second zonal-nonzonal interaction term, 
$\langle\mathbf{v}_{E,{\mathrm{Z}}}\rangle\cdot\nabla h_{\mathrm{NZ}}$, 
is of the same order as (\ref{ZNZint}) if we assume that  
\begin{equation}
\frac{h}{F} \sim \frac{e\phi}{T}
\label{h_phi_ordering}
\end{equation}
for both zonal and nonzonal modes. Balancing (\ref{ZNZint}) and (\ref{NZdrive}), 
we find, after cancellation of $k_y \phi_{\mathrm{NZ}}$, that 
\begin{equation}
k_{\mathrm{Z}} h_{\mathrm{Z}} \sim \frac{F}{L_T}.
\label{constraint_from_nz}
\end{equation}
This is a form of mixing-length hypothesis \cite{Wesson_mixing_length}, 
suggesting that the perturbed zonal gradients $\sim k_{\mathrm{Z}} h_{\mathrm{Z}}$ compensate
the background equilibrium gradients associated with $F$.\footnote{They need not necessarily flatten the background gradient completely or everywhere. From Figure \ref{zv_overlap}, the normalised zonal velocity $k_{\mathrm{Z}}\rho_e e\phi_{\mathrm{Z}}/T\rho_\ast \approx 0.3$ (we also find $k_{\mathrm{Z}}\rho_e\delta T_{\mathrm{Z}}/T\rho_\ast\approx 0.3$), which should be compared with the background gradient $a/L_T = 3.3$. For the (random-noise) initial conditions used in our simulations, the minimum $a/L_T$ required to sustain turbulence is between $3.0$ and $3.3$, so the perturbation levels correspond roughly to the distance away from this nonlinear critical gradient (of which we have not made a precise determination because the system appears to be strictly subcritical in the presence of even small flow shear; the linear critical gradient {\em without} flow shear is about $2.4$, as shown in \ref{linear_simulations}). Furthermore, the perturbed gradient is not constant (since $k_{\mathrm{Z}}\neq 0$); the actual magnitude of the gradient at any particular $x$ can exceed 0.3. 
}
It follows from (\ref{h_phi_ordering}) and (\ref{constraint_from_nz}) that
\begin{equation}
k_{\mathrm{Z}} \frac{e\phi_{\mathrm{Z}}}{T} \sim \frac{1}{L_T}.
\label{zv_equation}
\end{equation}
Thus, the gradients of the zonal modes (e.g., the zonal velocity or the zonal temperature 
gradient) are independent of collisionality. This independence is indeed seen in Figure \ref{zv_scaling}.

The only nonlinearity present in the zonal equation (\ref{GKE_Z}) is the nonzonal-nonzonal interaction
term (IV); zonal modes are not directly driven by background gradients 
because $\nabla F$ is in the $x$ direction. 
One can think of (\ref{GKE_Z}) as a kind of Langevin equation for zonal modes, which are
excited by coupling between nonzonal modes and damped by collisions (at long times, 
the only damping mechanism is collisional; see \ref{Felix}).\footnote{How precisely 
the zonal modes are generated is a matter for future research, but see section \ref{additional}.}  
Therefore, the dominant balance in (\ref{GKE_Z}) is between the nonlinear energy injection (IV) 
and collisional damping (III): 
\begin{equation}
\frac{c}{B} k_{\mathrm{Z}} k_y \phi_{\mathrm{NZ}} h_{\mathrm{NZ}} \sim \gamma_{\mathrm{Z}} h_{\mathrm{Z}},
\label{constraint_from_z}
\end{equation}
where $\gamma_{\mathrm{Z}}$ is the collisional damping rate of the zonal modes. 
Note that in conjecturing such a balance, we are assuming the zonal modes 
are not subject to a strong nonlinear instability that would break them down 
back into nonzonal perturbations, thus resulting in a purely collisionless 
saturated state. This possibility (which, for example, appears to be realised 
for zonal flows in certain regimes of ITG turbulence, where it is known as the 
tertiary instability \cite{Rogers00}) would amount to a dominant balance 
in (\ref{GKE_Z}) between the energy-injection and energy-removal 
effects within the nonlinear term (IV). We are expressly assuming that this is 
not the dominant balance in our near-marginal ETG turbulence.

Combining equations (\ref{zv_equation}), (\ref{constraint_from_z}), and (\ref{h_phi_ordering}) again, 
we get
\begin{equation}
\frac{h_{\mathrm{NZ}}^2}{h_{\mathrm{Z}}^2} \sim \frac{\phi_{\mathrm{NZ}}^2}{\phi_{\mathrm{Z}}^2} 
\sim \frac{\gamma_{\mathrm{Z}} eBL_T}{c k_y T} \sim \frac{1}{k_y\rho_e}\frac{\gamma_{\mathrm{Z}}}{v_{te}/L_T}.
\label{hratio_scaling_equation}
\end{equation}
Therefore, estimating the heat flux (\ref{Qdef}), we find 
\begin{equation}
\frac{Q}{\Qgb}
\sim 
\frac{n \delta T_\mathrm{NZ} v_{Ex}}{n T v_{te} \rho_\ast^2}
\sim
\frac{\delta T_{\mathrm{NZ}}}{T}\frac{c k_y \phi_{\mathrm{NZ}}}{B v_{te}\rho_\ast^2}
\sim
k_y\rho_e\left(\frac{e\phi_{\mathrm{NZ}}}{T \rho_\ast}\right)^2,
\label{Q_form}
\end{equation}
where we have used (\ref{h_phi_ordering}) to estimate ${\delta T}/{T}\sim {e \phi}/{T}$.
Using (\ref{hratio_scaling_equation}) to relate $\phi_{\mathrm{NZ}}^2$ to $\phi_{\mathrm{Z}}^2$ 
and (\ref{zv_equation}) to estimate $\phi_{\mathrm{Z}}^2$, we get 
\begin{equation}
\frac{Q}{\Qgb}
\sim 
\frac{\gamma_{\mathrm{Z}}}{v_{te}/L_T}\left(\frac{e\phi_{\mathrm{Z}}}{T \rho_\ast}\right)^2
\sim
\frac{\gamma_{\mathrm{Z}}}{k_{\mathrm{Z}}^2\rho_e^2(v_{te}/a)}\frac{a}{L_T}.
\label{conductivity}
\end{equation}
Thus, we are able to estimate the electron heat conductivity entirely in terms of the linear damping rate and characteristic scale of the zonal flows.
It is possible to show analytically (\ref{Felix}) and confirm numerically (section \ref{zonal_damping_section}) that in the long term, zonal modes are damped by collisional (Ohmic) resistivity:
\begin{equation}
\gamma_{\mathrm{Z}}\sim\nu_{ei}k_{\mathrm{Z}}^2\rho_{pe}^2,
\label{gammaZ}
\end{equation}
where $\nu_{ei}$ is the electron-ion collision rate and $\rho_{pe} = \rho_e B/B_p$ is the 
``poloidal Larmor radius'' of the electrons ($B_p$ is the poloidal magnetic field; see 
\ref{order1} for a more precise definition of $\rho_{pe}$). 
Using (\ref{gammaZ}) in (\ref{conductivity}), we finally obtain
\begin{equation}
\frac{Q}{\Qgb}\sim
\frac{\nu_{ei}}{v_{te}/a}\left(\frac{B}{B_p}\right)^2 \frac{a}{L_T}\propto\nu_{ei}.
\label{Q_scaling}
\end{equation}
Thus, a relatively simple theoretical argument 
has produced a linear scaling of the heat flux with collisionality. Note that all dependence on $k_\mathrm{Z}$ 
or any other wave numbers has cancelled in the final expression (\ref{Q_scaling}), 
and so, in order to obtain the heat flux, we need not know 
the spatial scales of either zonal or nonzonal modes. Considering the simplicity of the argument, and the level of agreement between it, our numerical results (Figure \ref{collscan}), and the experimental MAST scaling (\ref{MASTscaling}), we find it quite compelling.

\subsection{Damping of zonal modes}
\label{zonal_damping_section}

In the theoretical argument of section \ref{twiddle}, 
a crucial step was to use the expression (\ref{gammaZ}) for the 
collisional damping of the zonal modes, which allowed 
us to estimate the heat flux according to (\ref{conductivity}) and avoid having to 
theorise about the characteristic scale of the zonal modes (a nontrivial question, with, as our simulations indicate, possibly 
a nonuniversal answer). 
In \ref{Felix}, 
the damping rate (\ref{gammaZ}) is derived analytically. 
Physically, the situation can be summarised as follows. 

\begin{figure}[t]
\centering
\includegraphics[scale=0.8]{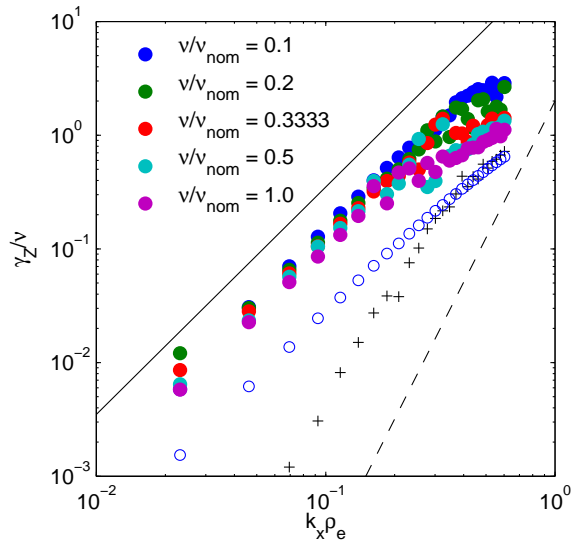} 
\caption{Zonal damping rate normalised to collisionality, $\gamma_{\mathrm{Z}}/\nu$, versus $k_x\rho_e$, spanning the range of collisionalities shown in Figure \ref{collscan} (solid colors). The final states of various
saturated nonlinear simulations were used as initial conditions, with the nonlinearity switched off. Also shown (black crosses) are the corresponding damping rates for a simulation at $\nu = \nu_{\mathrm{nom}}$ in which electron-ion
collisions were turned off (formally by setting $Z_\mathrm{eff}=0$; see section \ref{colls}); and a simulation at $\nu = \nu_{\mathrm{nom}}$ in which electron-ion collisions were retained but
magnetic drifts were turned off (blue open circles). The solid line is the slope 
$\propto k_x^2$, corresponding to the scaling (\ref{gammaZ}); 
the dashed line is $\propto k_x^4$, corresponding to the scaling expected 
when $\nu_{ei}=0$ (see \ref{intraspecies}).}
\label{zonal_damping}
\end{figure}

Consider a zonal perturbation with some perpendicular wave number $k_x$ 
satisfying (\ref{GKE_Z}) with zero right-hand side --- a linear 
equation. In the absence of collisions, this perturbation will decay quickly 
(on the time scale $\sim a/v_{te}$), but not 
to zero, leaving a finite residual zonal field \cite{RH}. With collisions present, 
after a period of a few collision times, which is still much shorter than the damping 
time, $\nu^{-1}\ll\gamma_\mathrm{Z}^{-1}\sim (\nu k_x^2\rho_e^2)^{-1}$ in the 
long-wavelength limit $k_x\rho_e\ll1$, it is possible to show that, to lowest 
order in $k_x\rho_e$, the remaining perturbation is a perturbed Maxwellian with 
a density (or, equivalently, $\phi$) and a temperature perturbation, both constant 
on each flux surface. These  
perturbations then decay diffusively due to perpendicular particle diffusion (equivalently, resistivity) 
arising from the collision operator. We already saw in section \ref{colls} that 
the gyrokinetic collision operator (\ref{model_e}) 
contains FLR terms that have the form of a spatial diffusion. 
These terms correspond to the displacement of gyrocentres by distances $\sim\rho_e$ 
due to collisions during Larmor rotation. Solving the ``zonal transport'' problem 
more carefully, one can show that collisions also displace the centres of banana 
(and corresponding passing) orbits by distances of order the poloidal gyroradius
$\rho_{pe} = (B/B_p)\rho_e$, which is larger. This leads to the damping rate (\ref{gammaZ}). 
It is proprtional to $\nu_{ei}$ (rather than $\nu_{ee}$, on which it depends weakly) 
because it is essentially the Ohmic resistive 
damping of electron currents (both parallel and perpendicular; see \ref{order1}), and 
it is due to electron-ion friction (cf.~\cite{Helander}).

The calculation of \ref{Felix}, where this is demonstrated more carefully, can 
be checked in our numerical simulations, to ascertain that it is indeed this effect 
that is responsible for the zonal damping. Figure \ref{zonal_damping} shows the 
zonal damping rate normalised by the collision frequency, $\gamma_\mathrm{Z}/\nu$, 
for a number of simulations in which the nonlinearity in equation (\ref{GKE_Z}) was
turned off (the right-hand side set to zero) and the zonal field allowed to decay linearly.
We see that, for a range of collisionalities $\nu$ and in a broad range of wave numbers 
$k_x\rho_e$, the scaling (\ref{gammaZ}), $\gamma_\mathrm{Z}/\nu \sim k_x^2\rho_{pe}^2$, 
is followed quite well. As a further successful test, we find that if we turn off
magnetic drifts ($\mathbf{v}_{B}\cdot\nabla h$ in equation (\ref{GKE_Z})), thus 
removing the banana orbits, the zonal damping rates drop by close to an order of
magnitude (blue open circles in Figure \ref{zonal_damping}). This is roughly consistent 
with a reduction of $\gamma_\mathrm{Z}$ by a factor of $(B/B_p)^2\approx (qR/r)^2\approx 18$
in our geometry, 
to $\gamma_\mathrm{Z}\sim \nu k_x^2\rho_e^2$, 
with the dominant diffusion due in this case to finite Larmor orbits, as explained above. 

Finally, we test the theoretical expectation that the dominant contribution to the 
damping of the electron zonal flows, which are also currents, is proportional 
specifically to the electron-ion collision frequency $\nu_{ei}$. Figure \ref{zonal_damping}
(black crosses) shows that when the electron-ion collisions are turned off, 
$\nu_{ei}=0$, leaving only the (momentum-conserving) electron-electron collisions $\nu_{ee}$,   
the damping rates drop dramatically and scale as $\gamma_\mathrm{Z} \sim \nu_{ee} k_x^4\rho_{pe}^4$, 
as indeed expected theoretically (see \ref{intraspecies}). 

Thus, the scaling (\ref{gammaZ}) and the theory that leads to it (\ref{Felix}) appear to be sound 
and successfully reproduced in our simulations.\footnote{It is perhaps worth pointing 
out that such an agreement is only possible in simulations that use a
sufficiently realistic electron gyrokinetic collision operator (see section \ref{colls}), 
an indispensable property being momentum conservation by the electron-electron 
collisions and a correct capturing of Ohmic resistivity by the electron-ion ones. 
See, however, section \ref{zeffscansection} for certain simplifications that are allowed.} 

\subsection{Numerical tests of the role of collisions and zonal modes}
\label{modified_dynamics}

With the theoretical argument presented in section \ref{twiddle} in mind, let us now build up the evidence that the long-time steady state of the ETG turbulence in our simulations is controlled by zonal-nonzonal interactions and by the electron-ion collisional damping of the zonal modes. First, we remark that whilst
linear simulations (see \ref{linear_simulations})
indicate that the linear instability growth rates may be comparable to the nominal (experimental) 
collisionality, 
the growth rates depend only weakly on collisionality. It does not seem plausible that such insensitive linear physics can explain the strong collisionality dependence of the nonlinearly saturated state. Clues
to the actual (nonlinear) origin of this dependence can be obtained via nonlinear simulations with
modified dynamics, as described below.

\begin{figure}[t]
\centering
\includegraphics[scale=0.9]{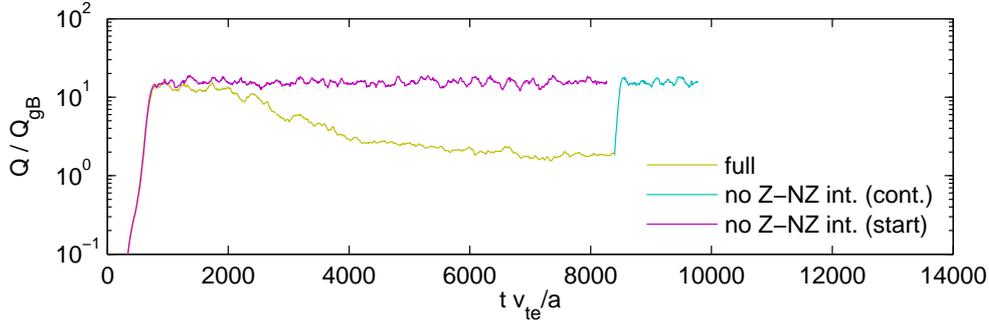} 
\caption{The large-box simulation shown in Figure \ref{overlap} (yellow) was restarted
at $t = 8397.7\, a/v_{te}$ 
without zonal interactions in the nonzonal evolution equation (cyan). The heat flux 
in this modified simulation returns to a level that is close to the high early ``quasi-saturated''
level. For direct comparison, the same modified simulation was also rerun from initial noise,
giving the same heat flux level (purple).}
\label{perts}
\end{figure}
First let us show that the zonal component regulates the amplitude of the rest of the turbulence, which
determines the heat flux. The cyan and purple curves in Figure \ref{perts} show the time evolution of the heat flux in simulations with identical parameters to one of the simulations in Figure \ref{overlap}, 
but with the nonlinear term artificially adjusted 
in such a way that the zonal modes no longer affect the evolution of the
nonzonal modes: namely, the zonal components have been zeroed out in the
calculation of the nonlinear term, so that in equation (\ref{GKE}), 
$\langle\mathbf{v}_{E}\rangle\cdot\nabla h$ is replaced by
$\langle\mathbf{v}_{E,\mathrm{NZ}}\rangle\cdot\nabla h_\mathrm{NZ}$. 
The zonal modes are still allowed to be nonlinearly driven by the nonzonal modes, 
but the zonal modes do not then feed back on the nonzonal modes; the
nonzonal evolution is entirely independent of the zonal evolution. 
This eliminates the nonlinear terms used to obtain the dominant balance (\ref{constraint_from_nz}). 
The heat-flux collapse occuring in the full simulation is prevented by this change, confirming that the collapse is indeed mediated by the effect of the zonal modes (which was turned off) on the nonzonal modes (which carry the heat flux).
The ``quasi-saturated'' streamer-dominated state is thus just the saturated state that would have emerged had the zonal flows been prohibited or suppressed.

\begin{figure}[t]
\centering
\includegraphics[scale=0.9]{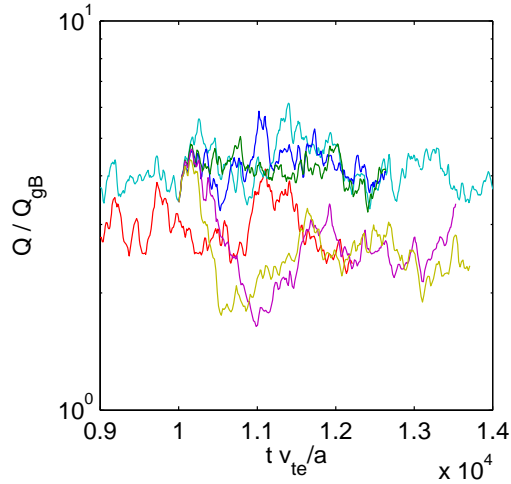} 
\caption{The two simulations corresponding to the red squares in Figure \ref{collscan} are shown: the large-box simulations with $a/L_T = 3.42$ and collisionalities $\nu = \nu_{\mathrm{nom}}/2$ (cyan) and $\nu = \nu_{\mathrm{nom}}/3$ (red).
The higher-collisionality case is restarted
at $t = 10003.6\, a/v_{te}$
with certain terms in the collision operator reduced
to the lower collisionality value, whilst others are retained at the higher value: (a) only electron-electron
collisions reduced (blue); (b) only electron-ion collisions reduced (purple); (c) only nonzonal collisions reduced (green); (d) only zonal collisions reduced (yellow).
}
\label{perts2}
\end{figure}
Collisions damp the zonal modes, and in this context are important precisely \emph{because} they are small, as this means that the finite zonal fields that emerge from any fast collisionless evolution \cite{RH} are damped
very weakly and so can grow to dynamically significant amplitudes and regulate the turbulence. 
Let us show that it is the electron-ion collisions that affect the zonal modes in the crucial way; 
these are momentum 
non-conserving (for electrons), as they act to relax the electron flow and thereby
dissipate the associated current --- in other words, they give rise to Ohmic resistivity
(see discussion in section \ref{colls}). 
In Figure \ref{perts2}, we show time evolution of the heat flux corresponding to two different collisionalities in the same series of simulations (the two shown as red squares in Figure \ref{collscan} --- the large-box simulations with $a/L_T = 3.42$). The heat flux in the higher-collisionality simulation is shown by the cyan curve, the heat flux in the lower-collisionality one by the red curve. If we rerun the higher-collisionality simulation with $\nu_{ei}$ unchanged but $\nu_{ee}$ reduced to the lower value (blue curve), or with the zonal collisionality
unchanged but the nonzonal collisionality reduced to the lower value (green curve), there is no
significant change in the saturated heat flux. 
By contrast, if we reduce only 
 $\nu_{ei}$, leaving $\nu_{ee}$ unchanged (purple curve), or if we reduce only the zonal collisionality, 
leaving the nonzonal collisionality unchanged (yellow curve), the heat flux drops to a value
consistent with the lower-collisionality case (red).
Thus it is the electron-ion collisions on the zonal component that primarily 
determine the heat-flux collisionality dependence. 

\subsection{Simplified simulations for extended collisionality range}
\label{zeffscansection}

\begin{figure}[t]
\centering
\subfigure[]{
  \includegraphics[scale=0.75,clip]{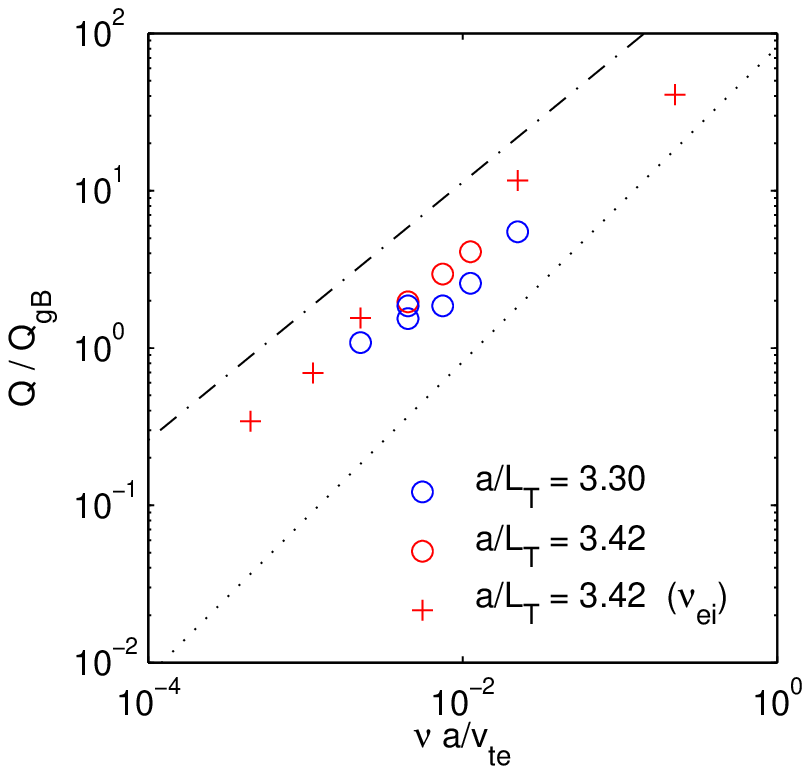} 
  \label{qe_scaling_zeff}
}
\subfigure[]{
  \includegraphics[scale=0.75,clip]{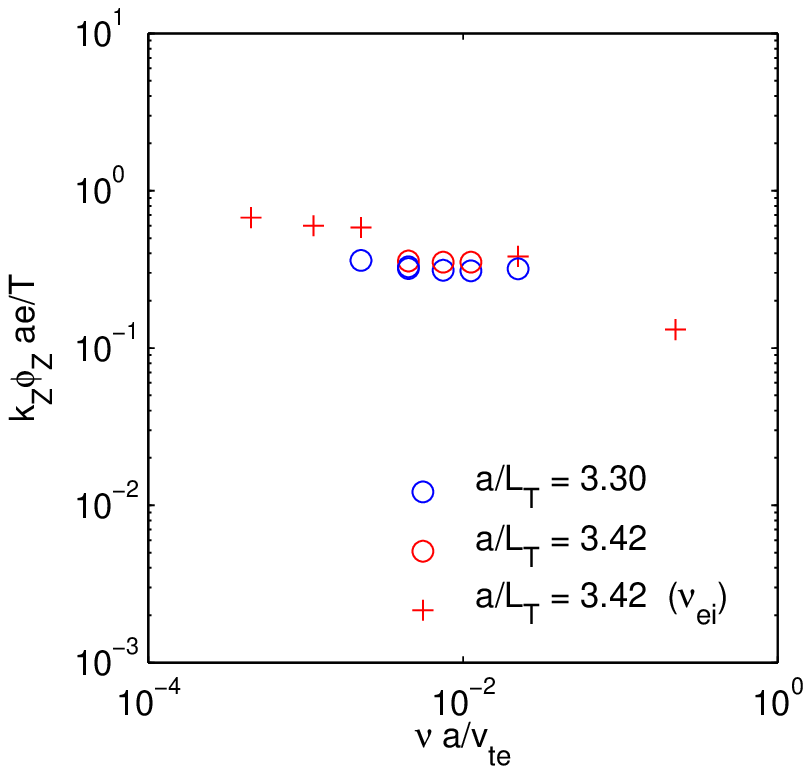} 
  \label{zv_scaling_zeff}
}
\caption{Variation of \subref{qe_scaling_zeff} the time-averaged electron heat flux $Q/\Qgb$, 
\subref{zv_scaling_zeff} the rms zonal velocity $k_\mathrm{Z} \phi_\mathrm{Z}$, 
adding to the points from Figure \ref{collscan} (now circles) further points (red crosses) obtained by varying only the electron-ion collisionality $\nu_{ei}$, but keeping the electron-electron collisionality at the nominal value, $\nu_{ee} = \nu_\mathrm{nom}$, all at the nominal temperature gradient $a/L_T = 3.42$. 
The dot-dashed line shows the experimental scaling (\ref{MASTscaling}) and 
the dotted line shows the theoretical linear scaling, $Q/\Qgb\propto\nu_{\ast}$, equation (\ref{Q_scaling}).}
\label{zeffscan}
\end{figure}

We have argued that it is the effect of electron-ion collisions on the zonal modes that sets the collisionality dependence of the saturated heat flux. Collisions in the nonzonal gyrokinetic equation (\ref{GKE_NZ}) are regularising, in that collisions are needed to dissipate the fine structure that the distribution function
develops in velocity space. This means that, while collisions cannot be dropped for the nonzonal modes, 
the particular value of $\nu$ is unimportant or, at most, has a weak effect --- for example 
on linear growth rates (see \ref{linear_simulations}).
This is similar to the well-known situation in fluid turbulence, where a small viscosity is needed to provide dissipation but the saturated state is independent of the exact value of this viscosity. This understanding of the underlying physics opens up an opportunity to probe the collisionality dependence of electron transport without paying the
high price of increased velocity-space resolution that reducing collisionality would exact. The
strategy is to vary only the collisionality affecting the zonal modes or only $\nu_{ei}$ 
(although in the latter case, the subdominant part of the zonal damping rate 
$\sim \nu_{ee} k_x^4\rho_e^4$ --- see section \ref{zonal_damping_section} and \ref{intraspecies} --- will 
eventually take over).\footnote{We further stress that one should be careful in the interpretation 
of such simplified simulations outside the range of collisionalities used for the full simulations. 
The simplified simulations could become unrepresentative of the original system if the fixed
electron-electron collisions suppress new modes that would otherwise have emerged and
dominated the dynamics at either low or high collisionality. 
We have not investigated whether this is the case for the simulations reported here.} 

Figure \ref{zeffscan} shows the collisionality dependence of the electron heat flux over a wider range of values of $\nu_{ei}$ than in Figure \ref{collscan}, accessed using the latter strategy: varying $\nu_{ei}$ only while 
keeping $\nu_{ee}$ at its nominal value. We see that the result of this extended collisionality scan is to
confirm the general plausibility of our picture of the ETG transport: $Q/\Qgb$ stays approximately proportional to $\nu_{ei}$ across a wider range of its values, and the independence of the
zonal flow velocity of $\nu_{ei}$ also approximately persists over this wider range. 
Note that these simulations were performed with
reduced perpendicular spatial resolution and without flow shear; 
see \ref{numerics} and \ref{convergence_study} for details. 

\subsection{Spatial structure of the saturated state}
\label{additional}

Finally, let us provide some details about the structure of the zonal and nonzonal components of the
turbulence in its final saturated state.

\begin{figure}[t]
\centering
\includegraphics[scale=0.65]{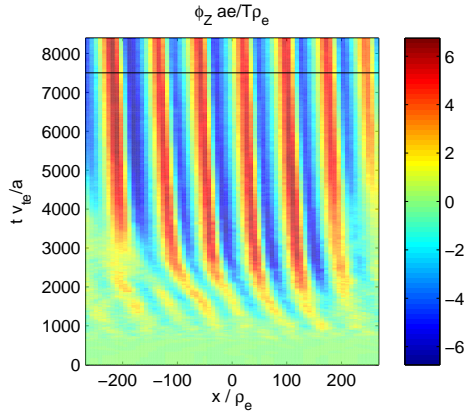} 
\caption{Zonal electrostatic potential $\phi_{\mathrm{Z}}$ in a flux-tube cross-section in the outboard midplane as a function of the radial spatial coordinate $x$ and time, for the same ``large-box'' simulation as shown in Figure \ref{overlap} ($\nu = 0.2 \,\nu_{\mathrm{nom}}$, $a/L_T = 3.3$). The black line shows the start of the time-averaging window used to characterise the saturated state.} 
\label{zp_hov}
\end{figure}

\begin{figure}[t]
\centering
\subfigure[]{
\includegraphics[scale=0.65]{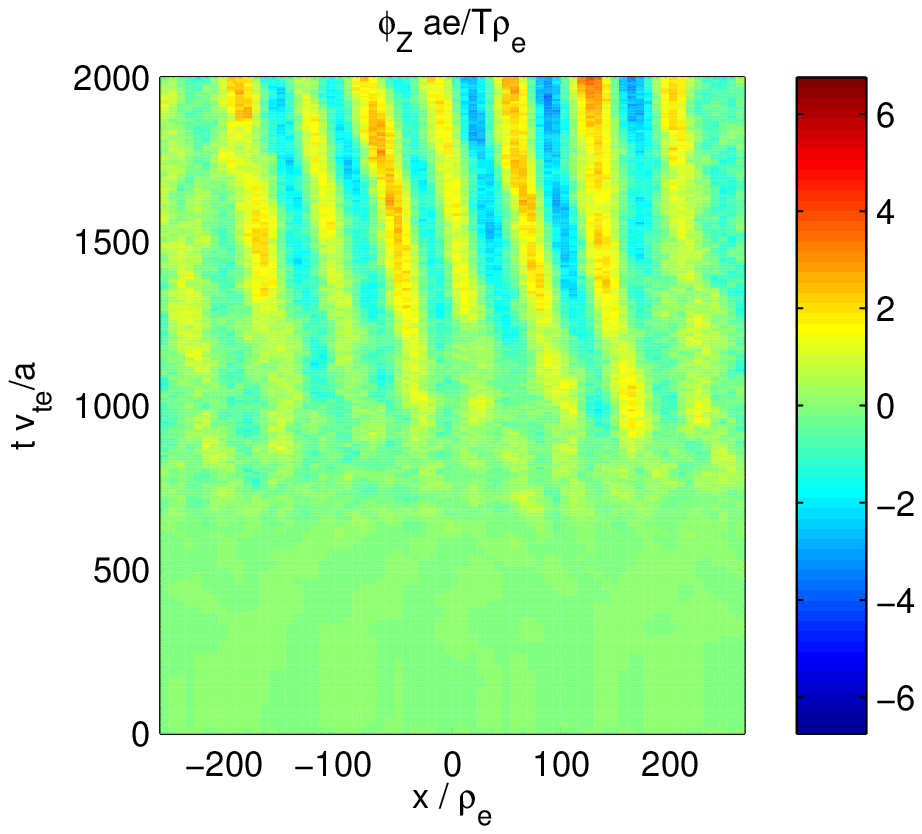}
  \label{zp_hov_start}
}
\subfigure[]{
\includegraphics[scale=0.65]{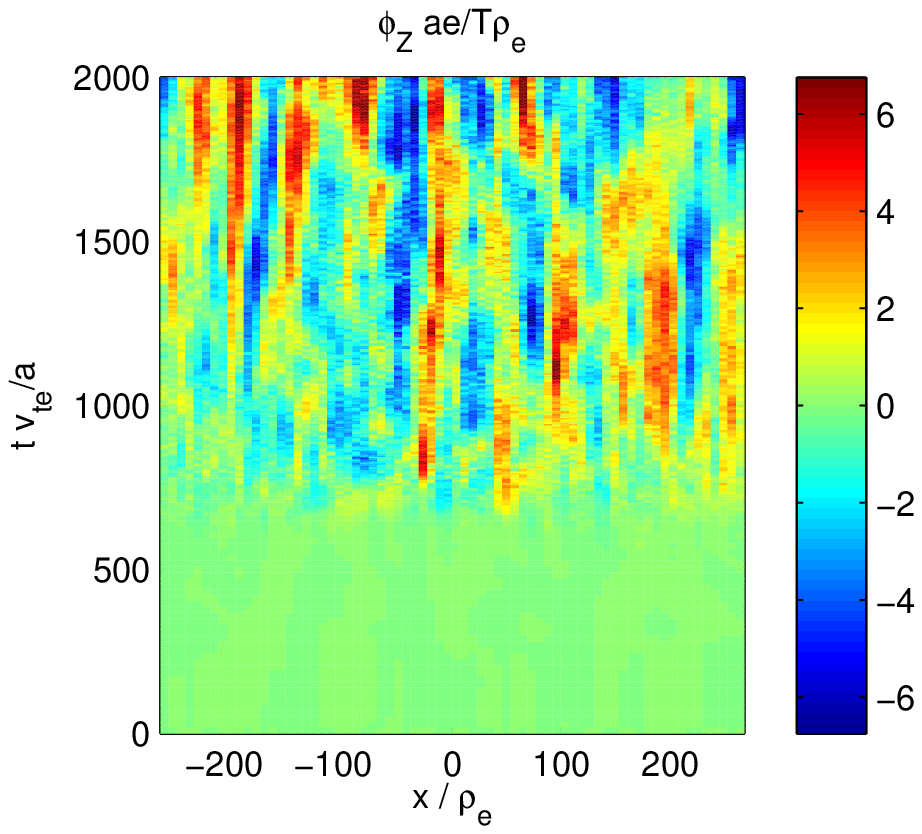}
  \label{zp_hov_start_zap}
}
\caption{
\subref{zp_hov_start} The same as Figure \ref{zp_hov}, 
but for the initial period $t\leqslant 2000 \,a/v_{te}$;
\subref{zp_hov_start_zap} the same for 
the case in which zonal feedback onto the nonzonal modes has been disconnected (the simulation 
for which the heat flux is shown as the purple curve in Figure~\ref{perts}).}
\label{zp_hov_start_comparison}
\end{figure}

The zonal component of the saturated turbulent field is long-lived, with a spatial structure that can persist, once established, for as long as the total simulation time. Figure \ref{zp_hov} is a Hovm\"oller (space-time) diagram of the zonal potential for the same case as shown in Figure \ref{overlap}. 
The zonal pattern barely changes during the time window that we have used
for time-averaging the heat flux and other quantities in the saturated state of this 
simulation.\footnote{In this context, we gain some insight into the mechanism 
of generation of the zonal modes by returning to the numerical experiment shown in Figure \ref{perts}
(section \ref{modified_dynamics}). We find that disconnecting the 
zonal feedback on the nonzonal modes but keeping the nonzonal-nonzonal interactions 
in the evolution equation for the zonal modes 
still leads to growth of the zonal modes in the ``quasi-saturated'' state, but they are quite different from 
the ones in simulations where their feedback is preserved: namely, 
they are more incoherent in time and they also grow more vigorously (see Figure \ref{zp_hov_start_comparison}). 
Thus, in the fully coupled system, even when the zonal modes are small, 
their excitation by the nonzonal modes still depends on the small modifications 
(subdominant as far as the heat flux is concerned) 
that they produce in the latter --- and thus the mechanism of this excitation  
is not as straightforward as just stochastic coupling of nonzonal modes into $k_y=0$.} 

\begin{figure}[t]
\centering
\subfigure[]{
  \includegraphics[scale=0.75]{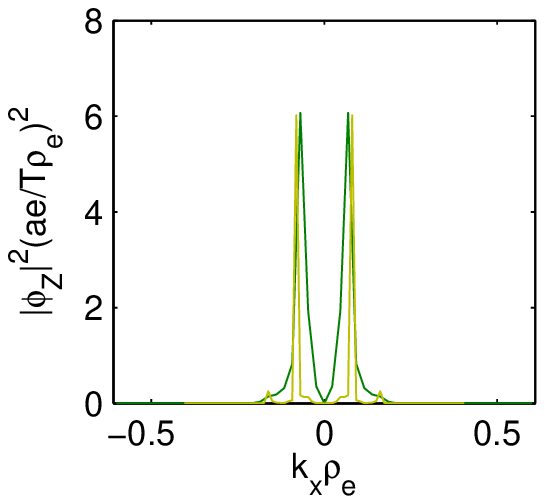} 
  \label{zonal_spectrum}
}
\subfigure[]{
  \includegraphics[scale=0.75]{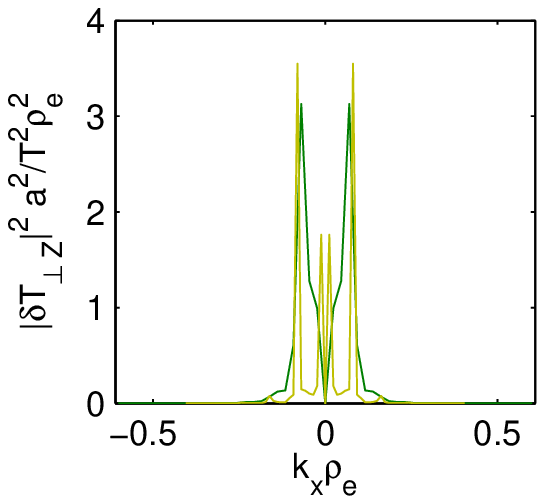} 
  \label{zonal_tperp_spectrum}
}
\subfigure[]{
  \includegraphics[scale=0.75]{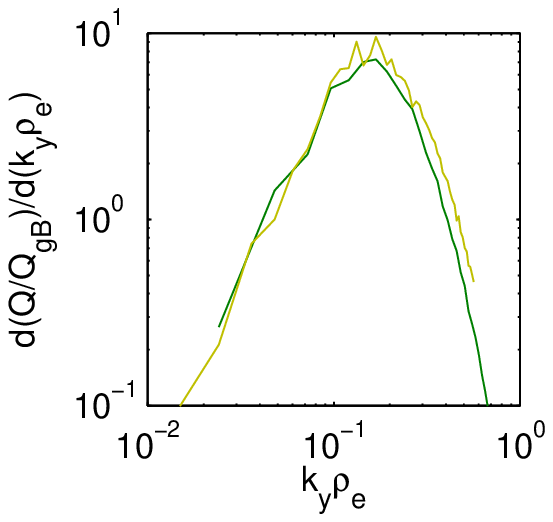}
  \label{kyspectrum}
}
\caption{For the same case as Figure \ref{overlap}: \subref{zonal_spectrum} the spectrum of the zonal potential; 
\subref{zonal_tperp_spectrum} the spectrum of the zonal perpendicular temperature perturbation; 
\subref{kyspectrum} the heat-flux spectrum; 
all are
averaged over the time window of the saturated state, which is shorter for the large-box simulation (yellow) than for the small-box simulation (green). 
}
\label{spectra}
\end{figure}

The spatial spectrum of the zonal potential is almost monochromatic in this particular case, as shown in Figure \ref{zonal_spectrum}. In general, we find that the zonal spectra typically have only a small number of sharp peaks. A single extra harmonic is just visible in Figure \ref{zonal_spectrum}.

\begin{figure}[t]
\centering
\subfigure[]{
  \includegraphics[scale=0.75]{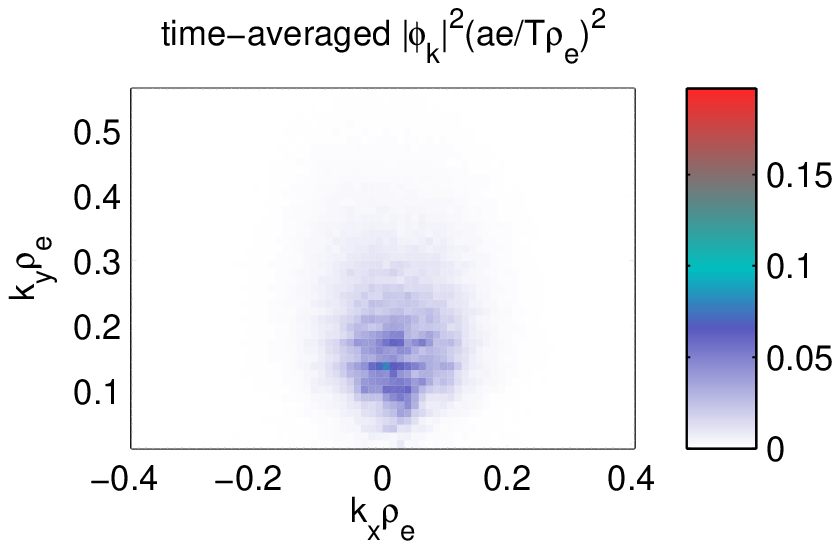}
  \label{meanspectrum}
}
\subfigure[]{
  \includegraphics[scale=0.75]{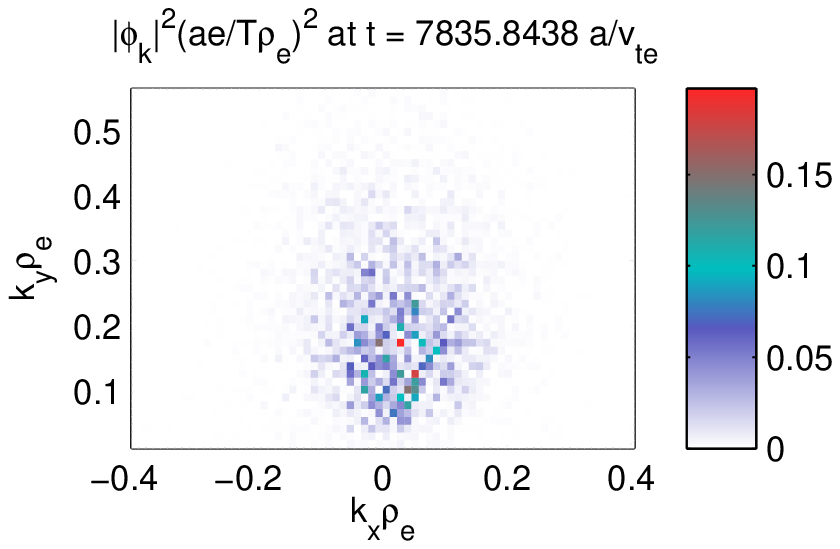}
  \label{slice1spectrum}
}
\subfigure[]{
  \includegraphics[scale=0.75]{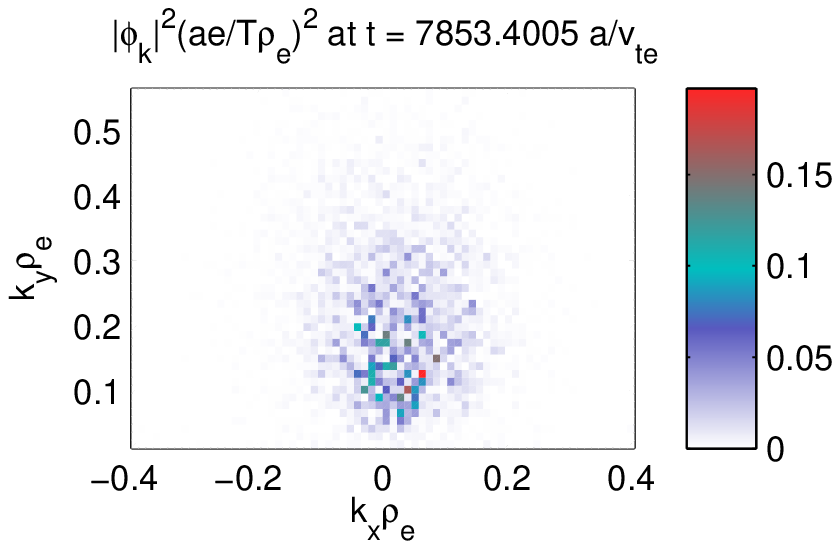}
  \label{slice2spectrum}
}
\subfigure[]{
  \includegraphics[scale=0.75]{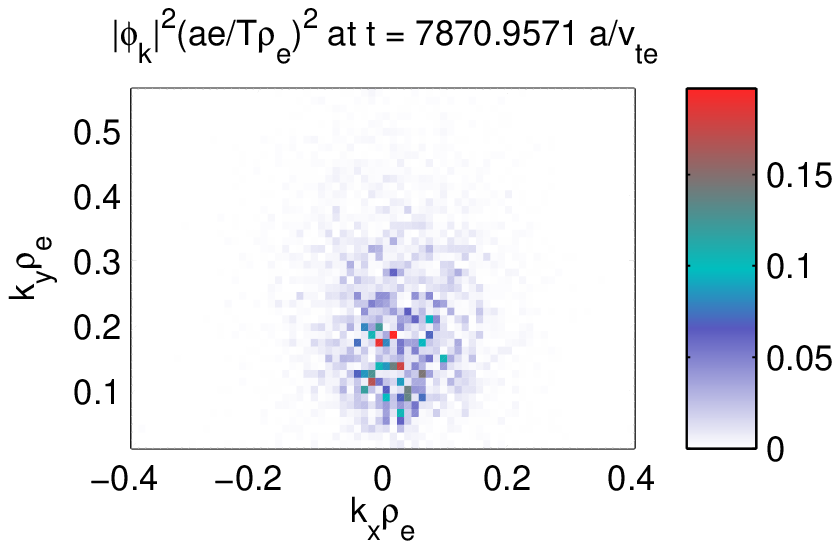}
  \label{slice3spectrum}
}
\caption{2D spectra of the nonzonal electrostatic potential for the same large-box simulation as in Figure \ref{overlap}: \subref{meanspectrum} the spectrum averaged over the saturated time-window $t \geqslant 7502.3 \, a/v_{te}$; \subref{slice1spectrum} instantaneous spectrum at the same time as Figure \ref{satslice}; \subref{slice2spectrum} \& \subref{slice3spectrum} instantaneous spectra at later times during the saturated state. The grid-cell size in these plots is 0.012 in both directions: the minimum positive $k_x\rho_e$ and  $k_y\rho_e$ (see \ref{numerics}).}
\label{flicker}
\end{figure}

By contrast with the monochromatic zonal spectrum, the nonzonal spectrum is broadband when averaged over time. Figure \ref{kyspectrum} shows
the heat-flux spectrum against $k_y$, and Figure \ref{meanspectrum} shows the nonzonal $\phi$ spectrum for the large-box simulation of Figure \ref{overlap} against both $k_x$ and $k_y$. 
Again by contrast with the static zonal component, the nonzonal component is
rapidly fluctuating. Figures \ref{flicker}\subref{slice1spectrum}-\subref{slice3spectrum} show the same 2D spectrum of $\phi$ at 
particular instants of time. Thus, the nonzonal spectrum has a ``flickering'' appearance: 
at any given time, a small number of modes are much more intense than the
others, but the dominant modes change over time, giving rise to a smooth time-averaged spectrum. This high ``$\mathbf k$-space intermittency'' is perhaps natural in the near-marginal saturated state at experimentally relevant parameters, with only a small number of modes excited at any given time. 
The details may be dependent on the simulation grid in $k_x$ and $k_y$, and this would be an interesting topic for future investigations.

\section{Summary and discussion}
\label{discussion}

The prevailing view of the structure of ETG turbulence in tokamaks and 
the associated levels of transport has its origin in the first gyrokinetic simulations,  
which did not include collisions \cite{Dorland00, Jenko00} (and did not correspond to near-threshold conditions in a spherical tokamak). 
Whereas the ITG turbulent state has long been believed to be zonal-flow dominated
\cite{Diamond05}, the ETG fluctuations were characterised by long radial eddies 
(``streamers'') that enhanced the transport to a level comparable with ITG turbulence, 
overcoming the reduction by a factor of $(m_e/m_i)^{1/2}$ expected from the relationship 
between the electron and ion gyroscales (the scales at which the two types of fluctuations 
were driven). 

The present study differs from the more traditional approach to modelling ETG turbulence 
in three respects: collisions are included; simulations are run for a 
much longer time; and the equilibrium parameters correspond to the experimental situation in a real device, namely MAST, and therefore place the system close to a marginal state with respect to the 
ETG drive (note also that MAST is a spherical tokamak, so has a somewhat different 
magnetic geometry compared to the more prevalent large fusion devices such as 
TFTR, JET or ITER). 

As a result, we find that, in application to the physical regime that we have considered, 
the standard picture of ETG transport is in need of revision. 
The high-transport, streamer-dominated nonlinear state does indeed emerge, and is not strongly
dependent on the collisionality of the plasma, but it persists only 
transiently, over relatively short simulation times (short compared to the energy 
confinement time but still long compared to a typical eddy turnover time, 
and long compared to typical simulation times used for ETG turbulence in the past). 
It turns out that this state 
is not entirely steady --- while the heat flux might appear to be statistically stationary, 
there is a slow growth of the zonal component of the fluctuations, 
which eventually (after $t v_{te}/a \sim$ a few 
thousand) reaches dynamical strength compared to the transport-setting nonzonal modes
and proceeds to change the character of the turbulence. A new, long-time,  
zonal-dominated saturated state emerges, whose structure is more reminiscent 
of what is traditionally expected of ITG, rather than ETG, turbulence 
(see Figure \ref{contours} and further discussion in section \ref{ITG}). 
We emphasize that it is the {\em final} saturation level of the heat flux in 
gyrokinetic flux-tube simulations, averaged over the turbulent fluctuation scales 
in length and time, that is physically relevant for transport calculations. 
We have found that the turbulent heat flux supported by the new long-time nonlinear 
state can be much lower than in the 
``quasi-saturated'' streamer-dominated state if the collisionality of the 
plasma is low (see Figures \ref{qe_overlap}, \ref{qe_scaling} and \ref{qe_scaling_zeff}).
The (roughly linear) collisionality dependence of the heat flux 
found in our simulations turns out to be in remarkably good agreement 
with the experimental scaling (\ref{MASTscaling}) \cite{Valovic}.\footnote[1]{Earlier 
simulations of ETG in MAST \cite{Joiner,Roach09} did report heat fluxes roughly consistent 
with the estimated experimental level of electron heat transport in the cases 
that were simulated. 
Whilst we obtain a difference in heat flux between the ``quasi-saturated'' state and the final state across the entire
range of collisionality shown in Figure \ref{collscan}, including at the nominal collisionality, 
the most dramatic differences occur when the collisionality is well below its nominal value (see Figure \ref{evolution}\subref{qe_evolution}).}

We have proposed a phenomenological argument (section \ref{twiddle}) 
whereby this collisionality scaling can be understood if one assumes that the saturation 
of the nonzonal modes is governed by the zonal gradients coming into approximate 
balance with the equilibrium gradients (to be more precise, 
the zonal-nonzonal interactions balancing the linear drive), while the saturation of the zonal modes 
is set by a balance between their nonlinear excitation by the nonzonal interactions and 
their damping by Ohmic resistivity. The latter is operative because electron-scale flows 
are also currents --- and so it is the electron-ion collisions that 
play the defining role in setting the zonal damping rate. We have supported our 
view by a series 
of numerical experiments that confirmed the crucial role of the zonal modes 
in enabling the emergence of the new saturated state (section \ref{modified_dynamics}) 
and the crucial role of the electron-ion collisions on the zonal modes 
in setting its collisionality dependence (sections \ref{modified_dynamics} and \ref{zeffscansection}).
The Ohmic damping of the zonal flows is an analytical result (\ref{Felix}), but 
we have also systematically confirmed that it is captured correctly in our simulations 
(section \ref{zonal_damping_section}). We have also documented some 
key qualitative features of the long-time saturated state: the long-time  
coherence and approximate monochromaticity of the zonal modes (cf.\ \cite{Parker_Krommes_2013,Parker_Krommes_2014}) and, in contrast, the dominant individual nonzonal modes ``flickering'' with time 
in and out of existence to give rise to a broad-band time-averaged spectrum 
(section \ref{additional}).   

\subsection{Previous work on ETG and the collisionality scaling} 

Long-time changes in the saturated state of gyrokinetic simulations have 
previously been reported in other numerical studies. 
Mantica et al.~\cite{ManticaEPS2011} reported a change 
in the zonal-nonzonal balance at long times in an ITG simulation. 
Guttenfelder and Candy~\cite{GuttenfelderCandy}, in their collisionless ETG simulations 
with adiabatic ions using the GYRO code, found a long-time reduction in transport 
associated with an increased level of zonal perturbations, which occurred at a low but 
not at a higher level of flow shear. 
We have not investigated here the dependence on flow shear, except via the simplified 
simulations with zero flow shear reported in section \ref{zeffscansection}, which appear to obey
the same scalings (Figure \ref{zeffscan}). 
It is not impossible that the transport scaling 
found in the present paper exists within a window in which the flow shear is not too large 
(although it should perhaps still be large enough to suppress ion-scale transport 
in order for a heat-flux calculation restricted to electron scales to make sense). 

Perhaps most relevantly for 
comparisons with our work, Nakata et al.~\cite{Nakata2010}, using 
a slab ETG model, adiabatic ions, a Lorentz collision operator (i.e., 
the pitch-angle scattering operator in (\ref{model_e})), and comparing two different 
sets of parameters, reported that transport at long times in their linearly more unstable case  
could be suppressed below the level of their linearly more stable case, owing to the formation
of zonal flows that collimated the turbulence into ``vortex streets'' and acted as a barrier 
to radial transport (see also earlier studies of ETG zonal flows \cite{Idomura05,Idomura06}). 
This suppression appears to be consistent with our findings. 

Recent direct measurements using the Doppler back-scattering system installed on MAST \cite{Hillesheim} are consistent with small-scale turbulence due to ETG.
Gurchenko and Gusakov \cite{GurchenkoGusakov} established an experimental correlation in several devices between electron-scale turbulence and anomalous electron transport, but not a causation pathway directly attributing the transport to ETG. The transport scaling that we obtain here derives from, and may, therefore, be an experimental signature for, not only the presence of electron-scale turbulence but also the saturation mechanism involving a dominant interaction with weakly damped zonal modes. In a regime in which this were not the saturation mechanism, the scaling of transport with collisionality may well be much weaker or absent.

An alternative explanation for the experimental collisionality scaling of heat
transport in spherical tokamaks that has been previously suggested relies on the transport 
associated with microtearing turbulence \cite{Valovic,Guttenfelder2012,Guttenfelder2013}. 
Since, in the present work, we have limited ourselves to electrostatic perturbations, 
the microtearing instability is excluded and and so it is clear that the collisionality 
scaling that we have found does not require it. 
Note that the arguments we apply to ETG-driven turbulence are fairly generic and 
may well apply in some form to other instabilities. We have, of course, also not excluded 
the possibility that other modes may produce a similar dependence for different reasons. 
The microtearing contribution in particular to the overall electron transport in real 
fusion devices remains a live and pressing research subject 
(made challenging, however, by the particular difficulty 
of obtaining well-resolved simulations of electromagnetic turbulence in tokamaks). 

It is an interesting question how general our picture 
of ETG turbulent transport might prove to be. NSTX \cite{Kaye} and MAST \cite{Valovic} both exhibit the strong scaling of confinement time with collisionality considered here, but in conventional tokamaks (as opposed to these two spherical ones), the scaling exponent between $B\tau_E$ and $\nu_\ast$ has been found to be much weaker or essentially zero \cite{ITERPB1999, Cordey2005NF, Bourdelle2011}.
Clearly, identifying the reason or reasons for this difference is an important issue for further study. 
We expect ETG to be more important in STs for the overall energy confinement, because of the large flow shear available to suppress ion-scale modes. Considering just ETG-dominated electron heat transport, however, one possibility is that, since in conventional tokamaks $B/B_p$ is large compared to STs, giving larger zonal damping, the heat flux given by the scaling (\ref{Q_scaling}) may be as large as the heat flux in our ``quasi-saturated''
state, in which case the ``quasi-saturated'' state may be the actual saturated state, not regulated by the mechanism discussed here, and with only weak or no collisionality dependence.\footnote{More generally, something similar may happen in ranges of collisionality, background gradients or other aspects of the magnetic geometry that differ from those considered here. However, even in conventional tokamaks, the collisionality scaling from the argument that we have presented may still be recovered at sufficiently low values of collisionality. Testing this hypothesis in simulations of conventional tokamaks would be an interesting avenue to explore in the future.}

\subsection{ETG vs.\ ITG turbulence near and far from threshold} 
\label{ITG}

The idea that the ETG turbulent state is dominated by zonal modes leads one naturally 
to the question of whether the saturated states of ETG and ITG turbulence are 
essentially similar, at least qualitatively. 

Our ETG turbulence model, equations (\ref{GKE}) and (\ref{QN2}), describes  
kinetic electrons with Boltzmann (adiabatic) ions.  
The simplest possible model for ITG turbulence would involve   
kinetic ions with Boltzmann electrons --- the latter physically justified 
by fast streaming of electrons along field lines. The main mathematical difference 
between these two models is that the Boltzmann electron response must be restricted 
to perturbations that have variation along magnetic-field lines. Namely,  
the density perturbation is, in contrast to the ETG case (\ref{ai}),  
\begin{equation}
\frac{\delta n}{n} = \frac{e(\phi-\bar\phi)}{T_e}, 
\label{ai2}
\end{equation}
where $\bar\phi$ is the flux-surface average \cite{Dorland93,Hammett93,AbelCowley}. 
It is this latter feature that is normally believed to be responsible for the 
difference between the zonal-flow-dominated ITG state and the streamer-dominated ETG state: 
radial variations in the zonal $\phi$ do not affect $\delta n$ as   
electrons cannot respond to radially varying zonal modes at ion scales; 
these can, therefore, grow large enough to break up the primary ITG-driven streamers 
via a secondary instability \cite{Cowley91,Rogers00} 
and isotropise the turbulence, whereas for ETG turbulence, 
the latter effect was believed to be too weak to destroy the streamers \cite{Dorland00,Jenko00} 
(although the physics that determined the radial scale 
of the streamers perhaps remained unclear \cite{Nevins06} --- possibly again a (weaker) secondary 
instability \cite{Cowley91,Jenko02}). 

What we have found in our simulations is that, whereas there is indeed no 
trace of a fast onset of a secondary instability similar to one that breaks 
up the ITG streamers (and is indeed seen in most ITG simulations), 
the ETG streamers' lease on life granted by a stronger density response 
to zonal perturbations is nevertheless only temporary: 
the zonal component does find a way to grow slowly, 
until it is large enough to take control over the nonzonal fluctuations.   

In addition to the ITG instability giving rise to streamer-like modes and 
the secondary instability breaking them up into zonal flows, the third pillar 
of the standard picture of ITG saturation is a ``tertiary instability'' 
whereby the zonal modes that have grown to a certain critical amplitude 
break up, returning energy to nonzonal perturbations \cite{Rogers00}. This mechanism 
of regulating the zonal component of the turbulence does not rely on collisional damping 
and thus should give rise to heat fluxes that are mostly independent 
of collisionality --- as indeed it appears to do, at least when the ITG 
turbulence is simulated in regimes that are far from the instability threshold \cite{Barnes11cb}. 
However, it has long been known that at temperature gradients that are close 
to the linear threshold (and thus, arguably, most relevant experimentally), 
the tertiary instability is ineffective, and a state of strongly suppressed 
transport ensues, called the Dimits upshift of the critical temperature 
gradient \cite{Dimits00,Rogers00}. Clearly, in the case of ETG turbulence, 
should a tertiary instability of the zonal modes appear, their collisional 
damping would cease to matter and consequently the collisionality dependence 
of the heat flux should flatten off (cf.\ \cite{Ricci06}). It is then tempting to suppose that 
our ETG state, in which the zonal modes are stable, collisionally damped 
and the heat flux scales with collisionality, is simply the ETG version 
of the Dimits-shift regime. Ascertaining whether this is indeed the case 
requires a larger parameter scan (in particular, in the temperature gradients) 
than has been undertaken here and has to be left for future work. 
By analogy with the ETG case, another tempting conjecture would be that the 
Dimits regime for the ITG turbulence should also have an (ion) collisionality 
scaling, by way of an argument similar to one presented 
in section \ref{twiddle}.\footnote{In porting this argument to the ion scales, 
one must not forget that an important difference between the ETG and ITG cases 
is the specially pivotal role played by electron-ion collisions for the ETG zonal modes. 
For ions, ion-electron collisions are always 
subdominant in the mass-ratio expansion because the ions carry nearly all of the momentum, 
and collisions with the lighter electrons have a negligible impact on the ion flow. 
By contrast, the relative flow of electrons (which constitutes a current) is strongly 
affected by collisions with the heavier, sluggish ions. Therefore, for electrons, collisions are 
not momentum-conserving, whereas for ions, to lowest approximation they are. 
How this difference plays out in the calculation of the rate of damping of ion 
zonal flows is outlined in \ref{ionzfs}. In the end, it turns out, however, 
that the damping rate scales in the same way as for the electron case, 
$\gamma_{\mathrm{Z}i}\sim \nu_{ii} k_x^2 \rho_{pi}^2$, although for a 
different reason.}
This in fact is not a new idea \cite{Lin,Lin00,Diamond05,Ricci06}, although the situation 
remains murky as including more realism (in particular, kinetic electron response)
in gyrokinetic simulations of ITG turbulent transport can 
nullify or even reverse its collisionality dependence \cite{MikkelsenDorland}. 
Since the collisionality scaling of the ETG transport that we report 
at least has some experimental support \cite{Valovic}, 
we feel more secure about the relevance of our conclusions --- although 
``stress-testing'' them by relaxing various assumptions (e.g., perhaps most urgently, 
allowing kinetic ions \cite{Nevins06}) and by extending the equilibrium parameter 
ranges is an essential direction for future research. 

Thus, the relationship between ETG and ITG is complicated and indeed within each 
of these two regimes, there are likely to be several sub-regimes depending 
on the part of the parameter space that one is interested in --- in particular, how 
far from the threshold the turbulence is. This said, our results suggest that there 
are perhaps more similarities between ETG and ITG physics than previously thought. 
Seeking a unified and universal description of various types of drift-wave turbulence 
may therefore be a worthwhile and reasonable aspiration. 

In conclusion, let us summarise again what has been achieved here. 
We have found that in the saturated state of ETG turbulence at driving gradients 
close to their experimental values in MAST, the electron heat transport decreases with 
collisionality, in agreement with experimental evidence \cite{Valovic}. 
This behaviour points toward improved confinement in future devices. 
We have explained it based on a simple theoretical picture (backed up by 
numerical tests) of an ETG 
turbulent state dominated by the interaction between nonzonal and zonal modes, 
with the collisionality dependence of the heat flux originating from the 
collisionality dependence of the resistive damping of the zonal modes. 
On a practical note, ETG simulations with generic initial conditions 
must be run to very long times to capture the (crucial) effect of zonal modes 
on the saturated state.

\ack{
This work was supported in part by EPSRC
under grant numbers 
EP/H002081/1, 
EP/I501045/1,  
EP/L000237/1 
and EP/M022463/1; 
and by the EC 
under grant agreement number 633053.
The work of AAS was also supported in part by grants from the UK STFC and EPSRC.
YcG was supported by the National R\&D Program through the National Research Foundation of Korea (grant number NRF-2014M1A7A1A01029835), and the KUSTAR-KAIST Institute.
WD was supported by the US DOE grants DE-FC02-08ER54964 and DE-FG02-93ER54197.
Computations were performed at the UK's HECToR and ARCHER services, on EFDA's HPC-FF facility, and on the Helios system at IFERC-CSC.
The authors thank
Ian Abel,
Jack Connor, 
Steve Cowley,
Greg Hammett,
Edmund Highcock,
Walter Guttenfelder
and
Ferdinand van Wyk 
for enlightening discussions; and also thank 
Justin Ball,
Francis Casson,
Paul Dellar,
Anthony Field,
Jon Hillesheim,
Nuno Loureiro,
Martin O'Brien,
Martin Valovi\v c
and two anonymous referees for helpful comments on versions of this paper. We are grateful to the Wolfgang Pauli Institute, Vienna, for its hospitality on several occasions.

}

\appendix
\section{Correspondence between \texorpdfstring{$B\tau_E$ and $Q/\Qgb$}{energy confinement time and heat flux} scalings}
\label{scaling_correspondence}

In the experimental scaling (\ref{MASTscaling}),
$\tau_E$ is the energy confinement time, i.e., the time it takes the power loss $P \sim QA$ to carry away the thermal energy $W \sim nTV$, where $A$ and $V$ scale with flux surface area and volume respectively, $Q$ is the energy flux, $n$ is density, and $T$ is temperature. Therefore,
\begin{equation}
\tau_E \sim \frac{W}{P} \sim \frac{nT a}{Q},
\label{tauE}
\end{equation}
where $a \sim V/A$ is the macroscopic length scale.

The appropriate normalising quantity for $Q$ is the ``gyro-Bohm'' heat flux $\Qgb = nTv_{te}\rho_\ast^2$, 
where $v_{te}=(2T/m)^{1/2}$ is the thermal speed, 
$\rho_\ast = \rho_e/a$, $\rho_e = v_{te}/\Omega_e$ is the gyroradius and $\Omega_e$ is the 
cyclotron frequency (since we are interested in electron transport, we are using electron 
quantities, but the argument for ions is exactly the same). 
In the gyrokinetic ordering, the heat flux is 
$Q = \mathcal O(\rho_\ast^2)$ automatically (see, e.g., \cite{Abel13}), 
so the normalised heat flux $Q/\Qgb$ must be order unity and independent of $\rho_\ast$. 
This normalised heat flux is
\begin{equation}
\frac{Q}{\Qgb} = \frac{Q}{n T v_{te} \rho_\ast^2} = \frac{Q}{n T a \Omega_e\rho_\ast^3} 
\sim \frac{1}{\Omega_e\tau_E \rho_\ast^3}, 
\label{QQgb}
\end{equation}
where we have used (\ref{tauE}) to obtain the last expression. 
At constant $\rho_\ast$, (\ref{QQgb}) implies 
\begin{equation}
\frac{Q}{\Qgb} \propto \frac{1}{B \tau_E},
\end{equation}
so we are probing the same dependence on collisionality by measuring $Q/\Qgb$ in 
gyrokinetic simulations as the experiments do by measuring $B\tau_E$ with $\rho_\ast$ fixed. 

Let us now explain what is meant when scaling (\ref{MASTscaling}) is claimed to hold 
experimentally \cite{Valovic}. The starting point is to assume that 
the following parametrisation of the energy confinement time holds: 
\begin{equation}
\Omega_e\tau_E = K\nu_\ast^{x_\nu}\rho_\ast^{x_\rho}q^{x_q} \beta^{x_\beta}\kappa^{x_\kappa},
\label{Ansatz}
\end{equation}
where $K$ is a constant, $\kappa$ is elongation \cite{Miller98}, 
$\beta$ is plasma beta, $q$ is the safety 
factor and
\begin{equation}
\nu_\ast = \frac{\nu}{\varepsilon^{3/2}v_{te}/qR}
\label{nustar}
\end{equation} 
is the effective electron collision rate $\nu/\varepsilon$ divided by the bounce 
frequency $\varepsilon^{1/2}v_{te}/qR$, where $\varepsilon = a/R$ is the inverse aspect ratio, 
and $R$ the major radius. This is the dimensionless quantity characterising collisional detrapping.
The experiments looking for the ``collisionality scaling'' 
strive to hold all these parameters constant apart from $\nu_\ast$ and measure 
the scaling exponent $x_\nu$ of $B\tau_E$ with respect to $\nu_\ast$.   
Since in our numerical study, we do not vary the equilibrium geometry, 
the same exponent $x_\nu$ will apply to the 
normalised collisionality in our gyrokinetic simulations, 
which is simply $\nu a/v_{te}$. 

Note finally that, in view of (\ref{QQgb}), if $Q/\Qgb$ is independent of $\rho_\ast$ 
(as it is hard-wired to be within the gyrokinetic ordering), it must be the case 
in (\ref{Ansatz}) that $x_\rho = -3$, which is indeed consistent with 
the experimental findings \cite{Valovic}.

\section{Plasma parameters}
\label{plasma_parameters}

The reference, or ``nominal'', plasma parameters for our simulations represent MAST shot 8500 at $t = 0.289$ s.
They are specified in Table \ref{parameters}. We use a
Miller \cite{Miller98} representation of the flux-surface geometry, which is assumed to be axisymmetric.

\begin{table}[htbp]
\center
\begin{tabular}{lcr} 
\hline
minor radius & $r/a$ & 0.65\phantom{0} \\
major radius & $R/a$ & 1.46\phantom{0} \\
safety factor & $q$ & 1.9\phantom{00} \\
magnetic shear & $\hat{s}$ & 1.8\phantom{00} \\
Shafranov shift & $R^\prime/a$ & $-0.25$\phantom{0} \\
pressure gradient & $p^\prime\beta/p$ & $-0.12$\phantom{0}\\
elongation & $\kappa$ & 1.57\phantom{0} \\
elongation gradient & $\kappa^\prime$ & 0.40\phantom{0} \\
triangularity & $\delta$ & 0.22\phantom{0} \\
triangularity gradient & $\delta^\prime$ & 0.16\phantom{0} \\
\hline
density gradient & $a/L_n$ & $-1.2\phantom{00}$ \\
electron temperature gradient & $a/L_T$ & 3.42\phantom{0} \\
effective ion charge & $Z_{\mathrm{eff}}$ & 1.15\phantom{0} \\
species temperature ratio & $T_e/T_i = 1/\tau$ & 1.06\phantom{0} \\
electron collisionality & $\nu_{\mathrm{nom}} a/v_{te}$ & 0.02\phantom{0}\\
flow shear & $\gamma_E a/v_{te}$ & $-0.003$\\
\hline
\end{tabular}
\caption{Nominal dimensionless local equilibrium parameters based on MAST shot 8500 at $t = 0.289$ s. 
Here $a = 0.55$\,m is the half diameter of the last closed flux surface at the elevation of the magnetic axis and $r$ is the same for the particular flux surface we have chosen. 
The prime symbol denotes the derivative with respect to normalised minor radius: $^\prime\equiv a(d/dr)$. Scale lengths are defined by $a/L_n\equiv-n^\prime/n$ (density) and $a/L_T\equiv-T^\prime/T$ (temperature).
}
\label{parameters}
\end{table}

Note that the reversed density gradient is not rare in MAST \cite{McConeThesis} and was found to be linearly stabilizing \cite{Garzotti}.

In our simulations, we varied the collisionality $\nu$ relative to its nominal value $\nu_{\mathrm{nom}}$
given in Table \ref{parameters}. 
We carried out such a collisionality scan for two values of $a/L_T$, the nominal one given 
in Table \ref{parameters} and a slightly smaller (slightly more marginal) $a/L_T = 3.3$. 
All other parameters were held fixed.

A small amount of flow shear $\gamma_E$, given in Table~\ref{parameters}, 
was included in the nonlinear simulations. As in \cite{Roach09}, this was found to assist 
convergence and is of the same order as the experimental level in MAST (see \ref{shear}). 
The sign of $\gamma_E$ is unimportant because the equilibrium is up-down symmetric \cite{Parra11}. 

\section{Numerical details}
\label{numerical_parameters}

\subsection{GS2 spatial coordinates}
\label{coords}

GS2 solves the gyrokinetic equation (\ref{GKE}) in a flux tube that winds its way 
toroidally and poloidally around the toroidal flux surface designated by a label $r$ 
(see Table \ref{parameters}), which is defined as its half diameter at the elevation 
of the magnetic axis of the tokamak. The position along the flux tube 
is determined by the poloidal angle $\theta$, which can be thought of as 
the coordinate parallel to the magnetic field. The flux-tube domain that we use 
in GS2 extends over $-\pi\le\theta\le\pi$, where 
$\theta = 0$ is the outboard midplane and $\theta = \pm\pi$ the inboard midplane.

If we demand that 
\begin{equation}
\mathbf{B} = \nabla \alpha \times \nabla \psi, 
\label{Clebsch}
\end{equation}
where $\psi$ is the poloidal magnetic flux and $\alpha$ is chosen to satisfy (\ref{Clebsch}), 
each field line is labelled by the pair of Clebsch coordinates $(\psi,\alpha)$.  
Then $(\psi, \alpha, \theta)$ can be used as the three independent (but not orthogonal!) 
curvilinear coordinates. Designating the central field line of a flux tube by 
$(\psi_0,\alpha_0)$, the GS2 coordinates $(x,y)$ transverse to the magnetic-field direction are 
related to $(\psi,\alpha)$ in the following manner \cite{BeerCowley,EdmundThesis}: 
\begin{eqnarray}
\label{defx}
x &= \frac{q(\psi_0)}{B_0 r(\psi_0)} (\psi-\psi_0),\\ 
y &= \frac{1}{B_0}\left.\frac{d\psi}{d r}\right|_{r=r(\psi_0)} (\alpha - \alpha_0),
\label{defy}
\end{eqnarray}
where the safety factor $q(\psi)$ and the radial flux-surface label $r(\psi)$ are both 
flux functions and $B_0$ is a reference (constant) magnetic field defined to be 
the toroidal magnetic field at the mid-radius of the flux surface containing our flux tube. 
The $x$ coordinate is ``radial'' in the sense that it is transverse 
to the flux surface and increases away from the magnetic axis. 
The second coordinate $y$ is often called ``binormal'' because it 
effectively measures distances perpendicular to the field line 
but within the flux surface (at constant $\psi$). 
In the outboard midplane, $(x,y)$ form an approximately orthogonal grid 
in the cross-section of the flux tube 
perpendicular to the magnetic field, with $x$ and $y$ approximately 
measuring true distances from the centre of the flux tube. 
Following the field line around the flux surface, the flux tube twists 
due to magnetic shear, with $\nabla x$, $\nabla y$ and $\nabla \theta$ 
generally not orthogonal to each other. 

GS2 is pseudospectral with respect to $(x, y)$ but not $\theta$; that is, except for the
evaluation of the nonlinear term, its data grid has the coordinates $(k_x, k_y) = \mathbf{k}_\perp$ and $\theta$. For the evaluation of the nonlinear term, the Fast Fourier Transform is used
to go from $(k_x, k_y)$ to $(x,y)$ 
(and then back again after the evaluation). In the context of the GS2 grids, the terms ``parallel'' and ``perpendicular'' refer to $\theta$ and to the other two coordinates (either $(x, y)$ or $(k_x, k_y)$ as appropriate), 
respectively.

\subsection{Numerical grids, resolution and boundary conditions}
\label{numerics}

In our largest nonlinear simulations, 
the spatial grid has 48 cells in $\theta$ and $108 \times 144$ points in $x \times y$, the range of positive perpendicular wavenumbers being 
$(k_x\rho_{e},k_y\rho_{e})\in[0.012,0.40]\times[0.012,0.57]$.
We refer to these as ``large-box simulations'' in the text.
The simulations referred to as ``small-box simulations'' are the same except for the perpendicular grid parameters: the box size in real space is halved, but extended to somewhat higher perpendicular wavenumbers (i.e., higher spatial resolution): $80 \times 108$ in $x \times y$, spanning the spectral range 
$(k_x\rho_{e},k_y\rho_{e})\in[0.023,0.60]\times[0.024,0.84]$.
The parameter regimes in which these two classes of simulations are useful 
are discussed in \ref{convergence_study}. 
The wavenumber ranges are intended to capture electron physics; 
the Boltzmann ion response ensures that we need not worry about where they lie in relation to ion scales.

Dealiasing is used for the nonlinear term: zero-padding with higher-wavenumber modes is introduced before the Fourier step into real space; any nonlinear interactions that wrap around unphysically in reciprocal space will be nullified when these modes are thrown away after the inverse Fourier step.
The number of $x \times y$ points indicated here (applicable only to nonlinear runs) is the same as the total number of modes including the dealiasing modes, but the ranges of wavenumbers given above do not include the dealiasing modes. All data shown here, including plots in real space, derive solely from the evolved, non-dealiasing modes: the additional dealiasing modes are used only transiently during the nonlinear part of each time step, and are not included in the code's output.

Twist-and-shift parallel boundary conditions \cite{BeerThesis, BeerCowley} link together some of the
$k_x$ modes at low $k_y$, whereas the
$k_y = 0$ modes are periodic in $\theta$. At the remaining (open) domain ends, the parallel boundary condition is $h = 0$ for incoming particles. The boundary conditions are automatically periodic in the perpendicular
directions $(x,y)$ by virtue of the spectral representation.

The velocity-space grid has 18 energies ${\cal E} = v^2/2$, and, for passing particles, 16 grid-points in $\lambda = \mu/\cal E$ for each sign of $v_\parallel = \pm(2 {\cal E} - 2\mu B)^{1/2}$, where $\mu= v_\perp^2/2B$ is the magnetic moment. 
Trapped particles are accommodated by additional $\lambda$ grid points, one for particles bouncing at each $\theta$.

Linear simulations (\ref{linear_simulations}) use the same setup as nonlinear ones, but evolve each perpendicular wavenumber independently.

For the ``simplified'' simulations in section \ref{zeffscansection}, in which only the electron-ion collisionality was varied, the grid sizes were: 24 cells in $\theta$; $40 \times 54$ points in $x \times y$, spanning the spectral range 
$(k_x\rho_{e},k_y\rho_{e})\in[0.023,0.30]\times[0.024,0.41]$;
12 energies $\cal E$; and 8 $\lambda$'s for passing particles in each direction. This version of the code applied the zero incoming boundary condition to
$g = \langle\delta f\rangle = h + e\langle\phi\rangle F/T$,
rather than to $h$.\footnote{
The option to use $h$ was added to the code during the course of the present work. The difference 
between the two options vanishes as $\phi\rightarrow0$ at $\theta\rightarrow\pm\infty$, but it has been found in linear simulations \cite{ColinGreg} that the use of $h$ leads to faster convergence as the actual, finite
$\theta$ domain size is increased.}

\section{Linear growth rates}
\label{linear_simulations}

\begin{figure}[t]
\centering
\includegraphics[scale=0.75]{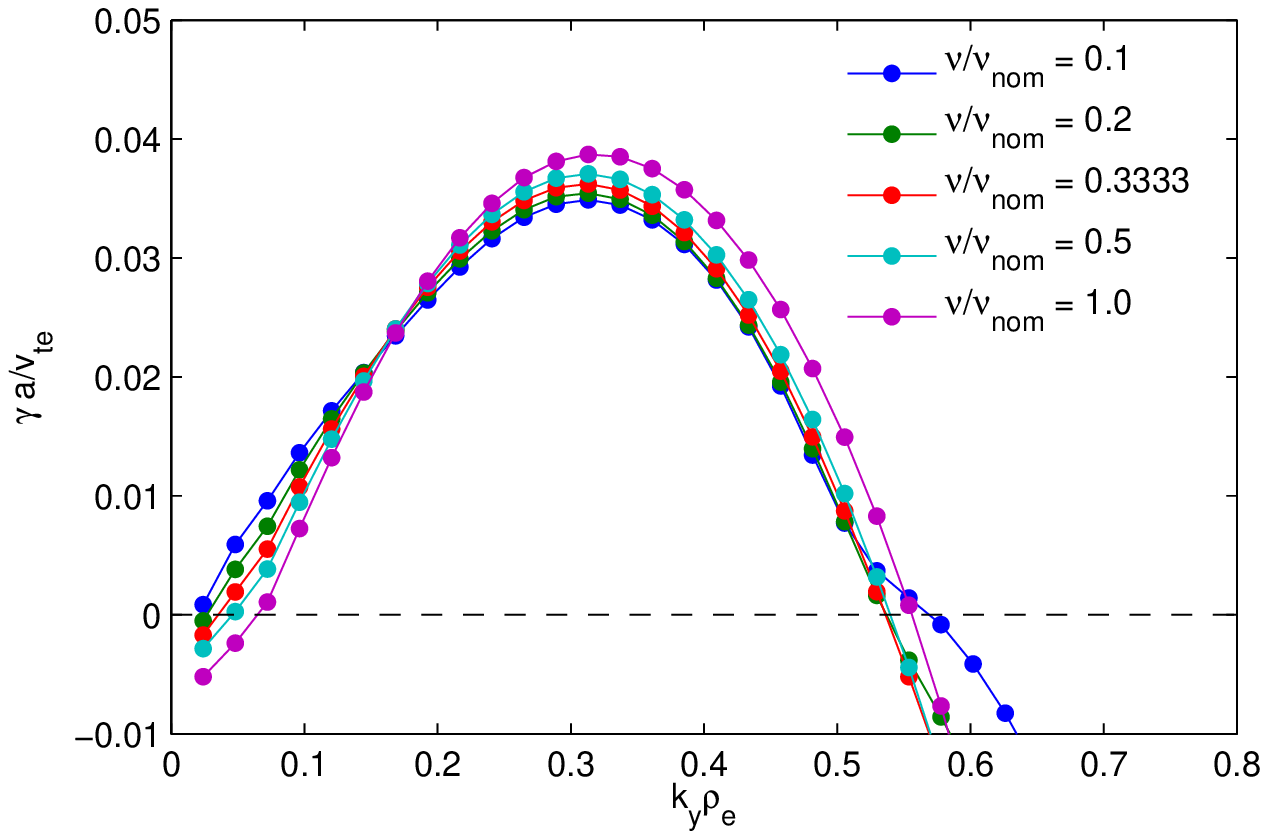} 
\caption{Linear growth rates versus $k_y$, at $k_x = 0$, with zero flow shear $\gamma_E = 0$,
for various collisionalities as indicated. 
All other parameters are the nominal ones given in Table \ref{parameters} (including $a/L_T=3.42$).
}
\label{linear}
\end{figure}

Figure \ref{linear} shows the growth rate of the most unstable linear mode at $k_x = 0$ 
for each $k_y$ in the numerical domain (which was described in \ref{numerics}). 
These linear simulations have zero flow shear, $\gamma_E=0$. 

At the given $a/L_T$, the effect of decreasing collisionality is to decrease slightly the maximum growth rate, and to extend to somewhat lower $k_y$ the unstable wavenumber interval. At 
$k_y\rho_{te}\approx 0.2$
where the saturated heat flux is maximal (see Figure \ref{kyspectrum}), the dependence of the linear spectrum on collisionality is weak.

Note that this scan is consistent with the previous observation \cite{Roach09} that
collisions lead to a stability gap between ETG- and ITG-driven modes (the range of $k_y$'s unstable to ITG being lower than the wave numbers shown here).

\begin{figure}[b] 
\centering
\includegraphics[scale=0.75]{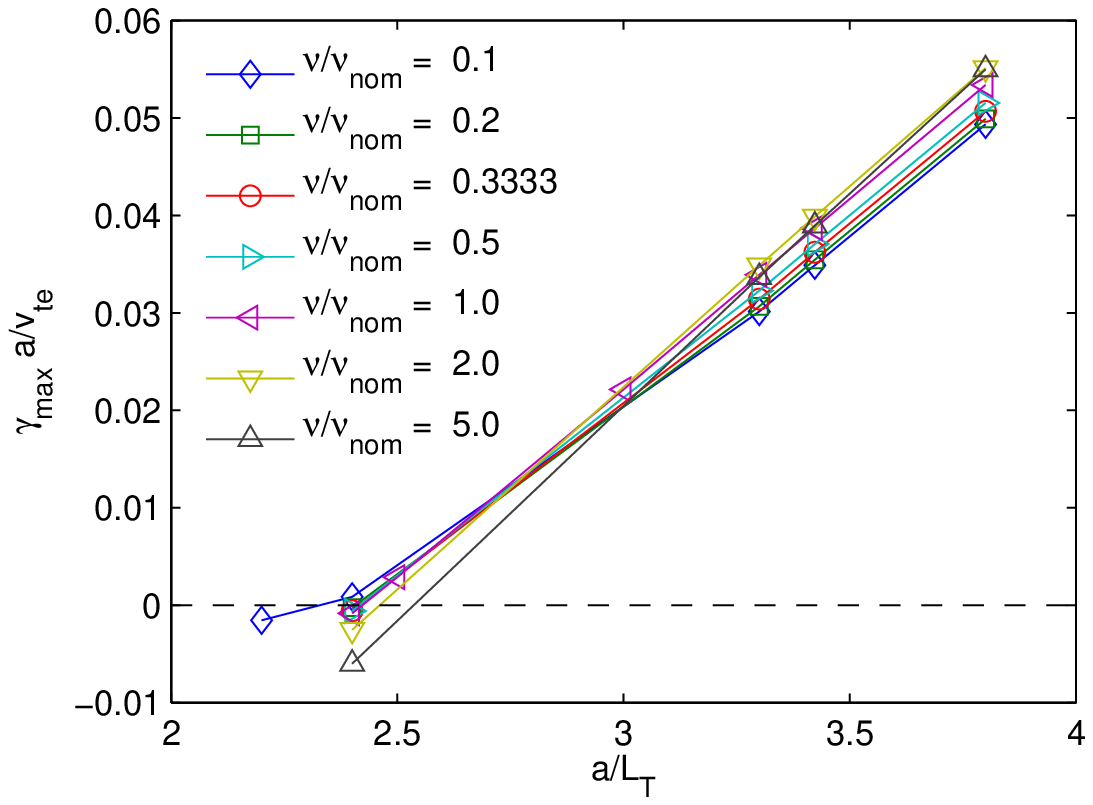} 
\caption{
Linear growth rates versus $a/L_T$, with zero flow shear $\gamma_E = 0$,
for various collisionalities as indicated, 
showing that the linear critical gradient (where the growth rate crosses zero) 
is insensitive to collisionality.
}
\label{critical}
\end{figure}

Figure \ref{critical} shows the dependence of the maximum growth rate on the temperature gradient 
(still with $\gamma_E=0$). We see that the linear critical gradient varies only between
$2.3$ and $2.5$ throughout this whole range spanning a factor of 50 in collisionality.
This is in agreement with the earlier findings \cite{Jenko2001} that collisions have very little
effect on the linear critical gradient.

Note that while these linear simulations were done at zero flow shear, our nonlinear runs 
had $\gamma_E = -0.003\,v_{te}/a$, a value representing the measured rotational shear in the MAST shot on which 
the rest of our equilibrium parameters were based. The ETG instability in the 
presence of flow shear can become subcritical in the sense that growth is transient 
and all modes eventually decay (cf.\ \cite{Newton10,Highcock10,Barnes11,Highcock11,Sch12,vanWyk16}). 
Given a finite initial perturbation, such transient growth still leads to a saturated nonlinear state
(because long enough transient growth can be as good as a formally unstable mode as far as 
maintaining turbulence is concerned \cite{Highcock10,Barnes11,Highcock11,vanWyk16}), 
although we have found the minimum value of $a/L_T$ required for this to be higher 
than the critical gradient implied by Figure \ref{critical}: it is typically between 
$3$ and $3.3$. The mapping out of a ``zero-turbulence manifold'' 
(cf.\ \cite{Highcock12,Ghim14}) for the ETG problem has been left outside the scope 
of this study --- and would be a formidable computational challenge, requiring very many 
very-long-time simulations. It appears plausible that the key parameters in such a study 
would be the electron temperature gradient $a/L_T$ and collisionality $\nu_{ei}a/v_{te}$, rather 
than the ion temperature gradient and the perpendicular flow shear 
$\gamma_E$, which has a more profound effect on the turbulence threshold in the ITG problem \cite{Highcock12,Ghim14,vanWyk16}. The magnetic geometry factor $qR/r\approx B/B_p$, which appears in the scaling (\ref{Q_scaling}), may play a similar role --- improving confinement as this quantity is reduced --- in both cases, although for rather different reasons.

\section{Issues of numerical convergence in nonlinear simulations}
\label{convergence_study}

We discuss here the selection of runs used to construct Figure \ref{collscan}, and the associated numerical convergence issues. We also include additional evolution plots, complementing the two that were shown in Figure \ref{overlap}. This appendix concludes with a short discussion on the role of flow shear.

\begin{figure}[t]
\centering
\subfigure[]{
  \includegraphics[scale=0.75,clip]{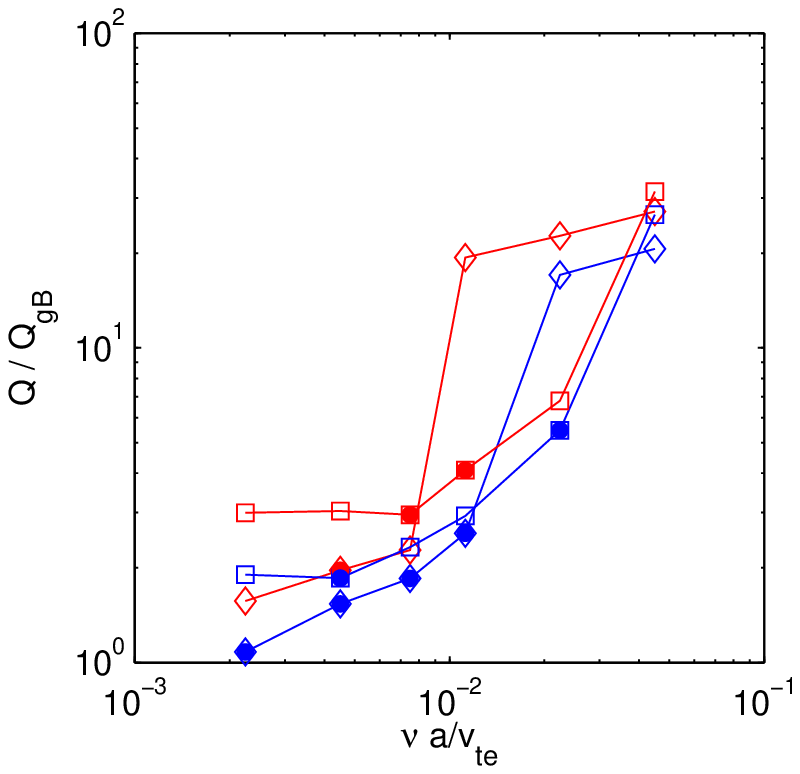} 
  \label{qe_uncvgd}
}
\subfigure[]{
  \includegraphics[scale=0.75,clip]{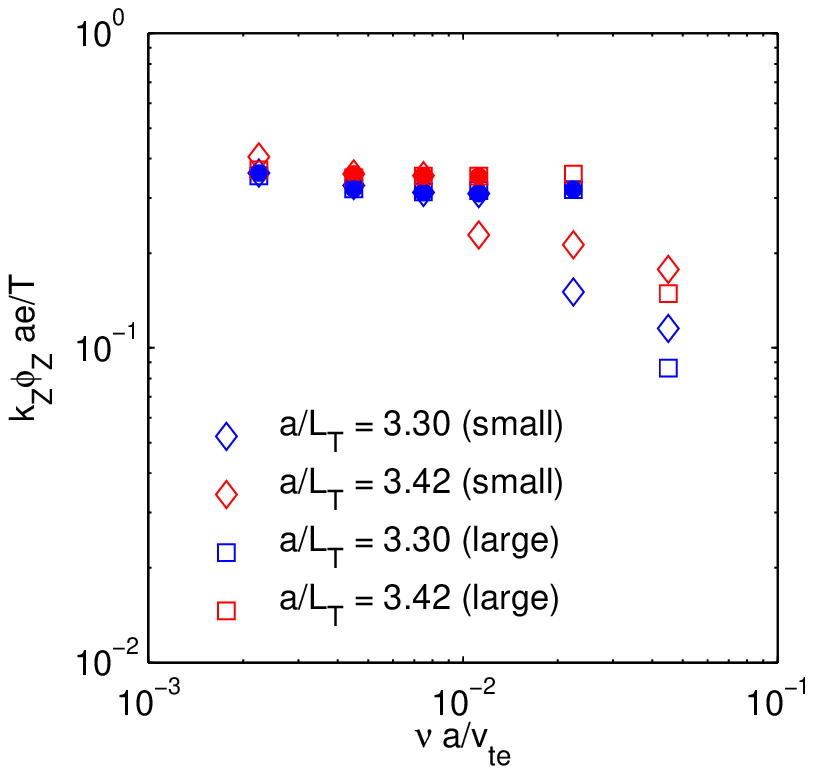} 
  \label{zv_uncvgd}
}
\caption{Variation of \subref{qe_uncvgd} the time-averaged normalised electron heat flux $Q/\Qgb$ and \subref{zv_uncvgd} the rms zonal velocity $k_\mathrm{Z} \phi_\mathrm{Z}$, defined by (\ref{kzphiz}), versus normalised electron collisionality $\nu a/v_{te}$ for our entire simulation
series using both small and large boxes, and two temperature gradients, including unresolved runs (hollow symbols; infilled symbols are the resolved points shown in Figure \ref{collscan}).}
\label{qescan_uncvgd}
\end{figure}

\begin{figure}[p]
\subfigure[]{ 
\includegraphics[scale=0.75]{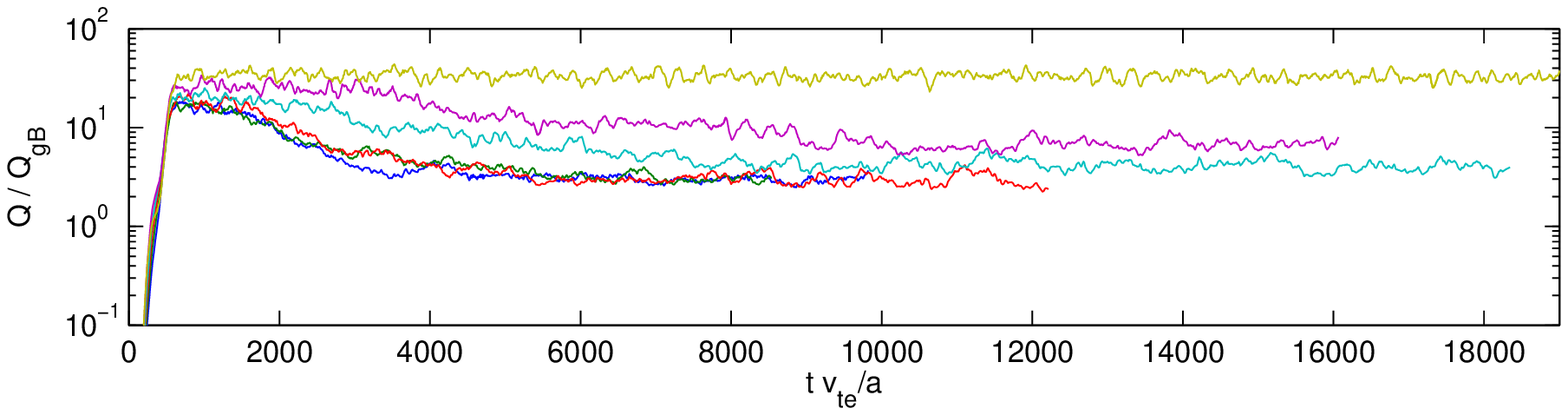} 
\label{qe_evolution}
} 
\subfigure[]{
\includegraphics[scale=0.75]{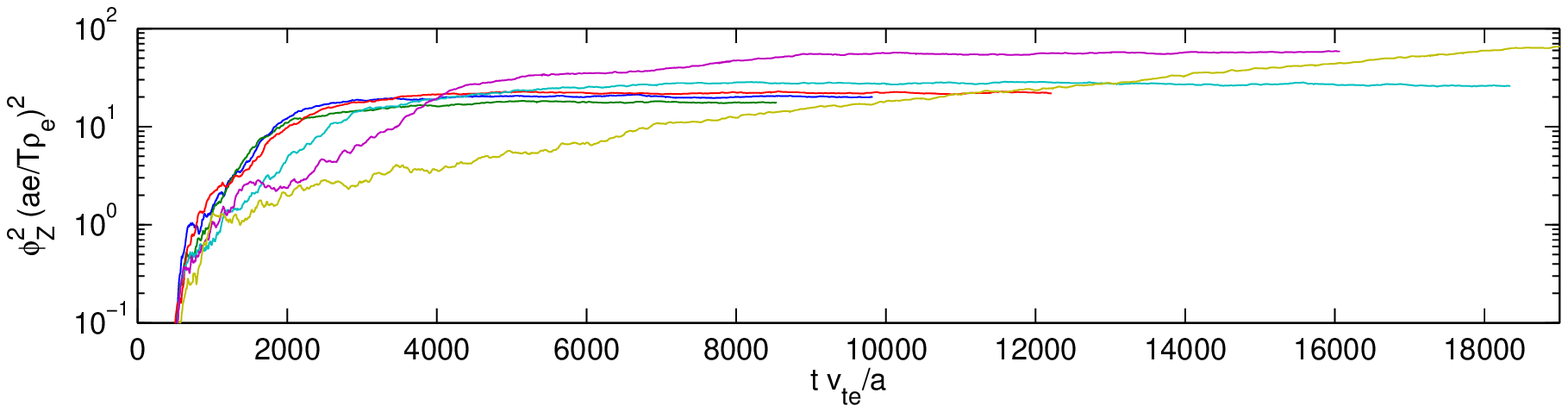}
\label{zp_evolution}
}
\subfigure[]{
\includegraphics[scale=0.75]{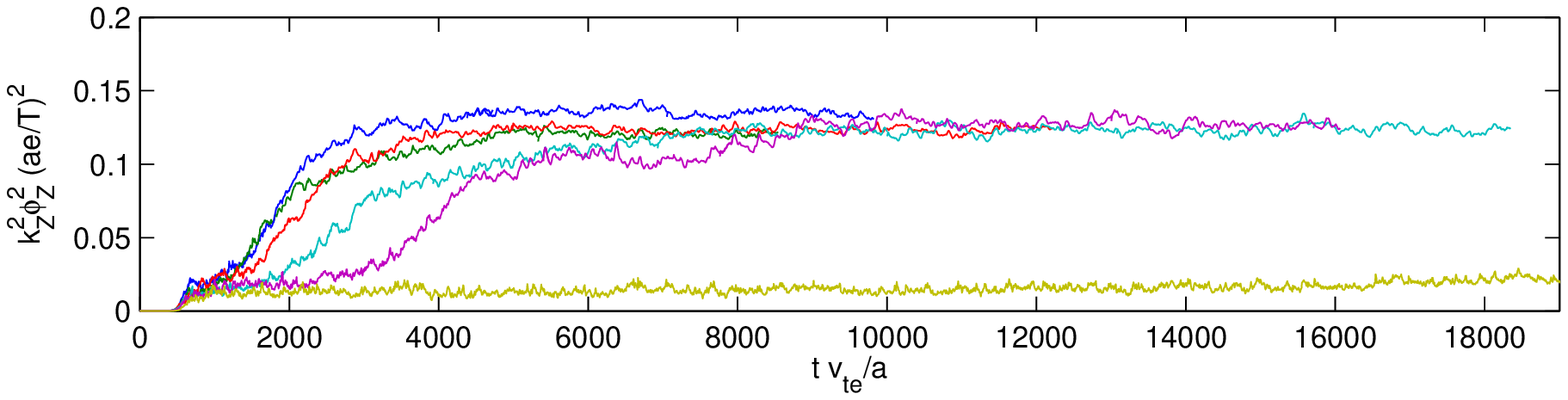}
\label{zv_evolution}
}
\subfigure[]{
\includegraphics[scale=0.75]{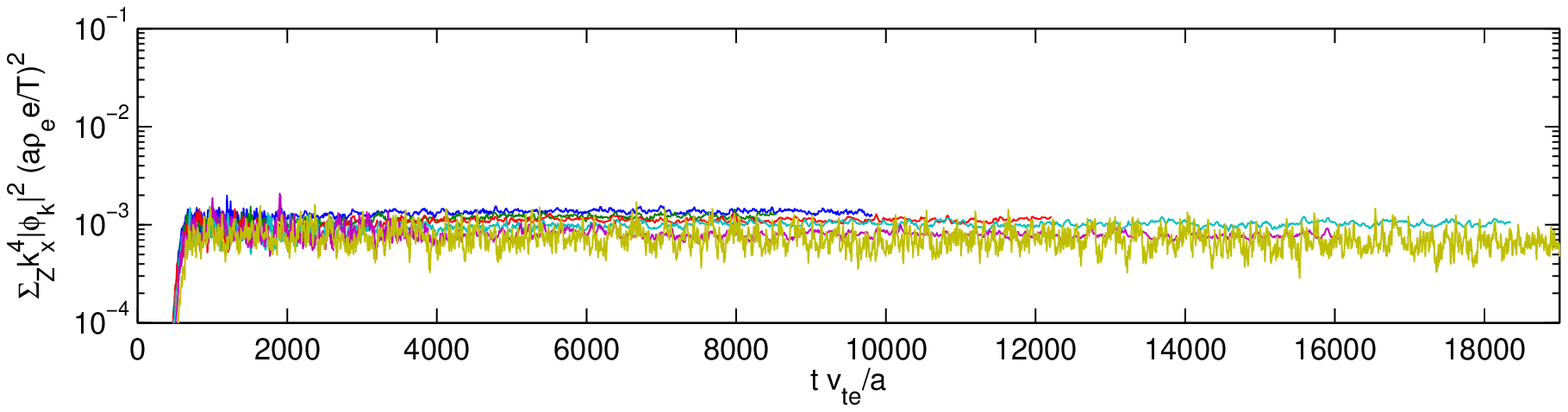}
\label{zs_evolution}
} 
\subfigure[]{
\includegraphics[scale=0.75]{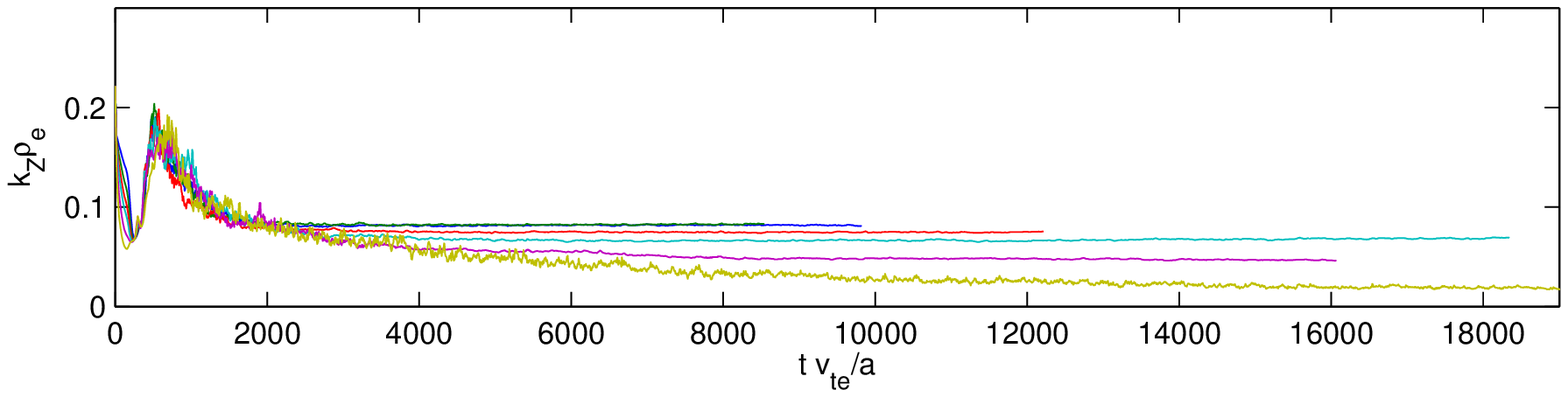}
\label{kz_evolution}
}
\caption{Evolution in time of \subref{qe_evolution} the normalised turbulent electron heat flux $Q/\Qgb$, \subref{zp_evolution} the square of the zonal electrostatic potential, $\phi_{\mathrm Z}^2 = \sum_{k_x} |\phi_{k_x,0}|^2$, \subref{zv_evolution} the square of the zonal velocity, $(k_{\mathrm Z}\phi_{\mathrm Z})^2 = \sum_{k_x} k_x^2|\phi_{k_x,0}|^2$, \subref{zs_evolution} the square of the zonal shear, $\sum_{k_x} k_x^4|\phi_{k_x,0}|^2$, and \subref{kz_evolution} the rms zonal wavenumber $k_{\mathrm Z}$, for large-box
simulations at the nominal $a/L_T = 3.42$ and at various collisionalities (colours: same as Figure \ref{linear}, plus yellow for $\nu = 2\,\nu_{\mathrm{nom}}$). 
} 
\label{evolution}
\end{figure}

\subsection{Convergence of the heat flux}

Figure \ref{collscan} was based on a set of simulations picked from a larger parameter scan that we have carried out. The choice of simulations was based on whether we deemed them to be numerically converged. Here we explain how this was decided.

Figure \ref{qescan_uncvgd} shows the results of both small- and large-box simulations 
(see \ref{numerics} for the explanation of what this means), over a range of collisionalities and for two values of the temperature gradient. Runs
excluded from Figure \ref{collscan} are shown as hollow points. They have been excluded because one
or more important physical quantity measured in these runs was found to be dominated by either the box scale (the lowest $k$) or by grid scales
(the highest $k$'s), and, therefore, the convergence of the simulation with respect to box size and/or spatial resolution could not be relied upon. 

Two particular pathologies have been identified, and are also visible in Figure \ref{evolution}, which shows the time evolution of the full set of large-box simulations at the nominal temperature gradient ($a/L_T=3.42$). 

(i) As the collisionality is reduced, it can be seen from the growth of the zonal velocity (Figure \ref{evolution}\subref{zv_evolution}) that the increasing zonal $\phi^2$ (Figure \ref{evolution}\subref{zp_evolution}) more than compensates 
initially for the gradually falling zonal wavenumber (Figure \ref{evolution}\subref{kz_evolution}), until the final level of zonal velocity is reached. However, the zonal shear (Figure \ref{evolution}\subref{zs_evolution}) remains relatively constant in time.
For the highest collisionality (yellow), the heat flux
never collapses, and we believe that in this case it is because the required zonal wavenumber
is too close to the simulation box scale; see Figure \ref{evolution}\subref{kz_evolution}. 
Such uncollapsed runs are visible as the high points in Figure \ref{qescan_uncvgd}\subref{qe_uncvgd}. 

\begin{figure}[t]
\centering
\subfigure[]{
\includegraphics[scale=0.75]{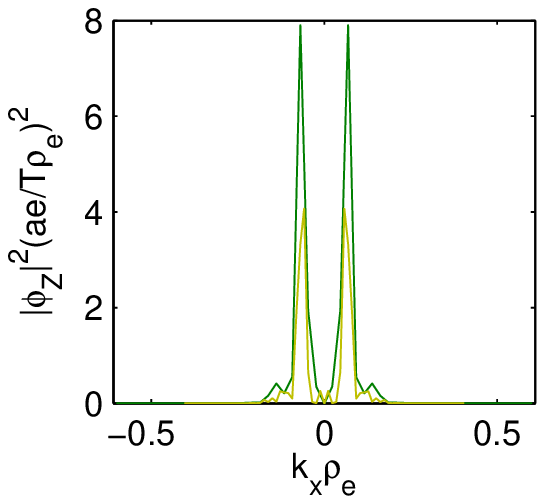} 
\label{uncvgd_zonal_spectrum}
}
\subfigure[]{
\includegraphics[scale=0.75]{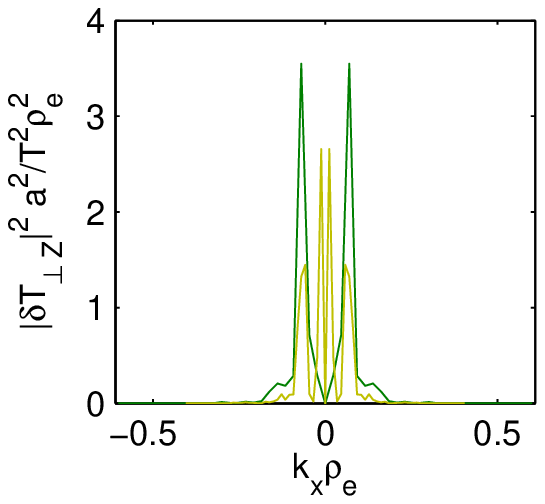} 
\label{uncvgd_zonal_tperp_spectrum}
}
\caption{Spectrum of \subref{uncvgd_zonal_spectrum} the zonal potential, \subref{uncvgd_zonal_tperp_spectrum} the zonal perpendicular temperature perturbation, for $\nu = 0.1 \,\nu_{\mathrm{nom}}$, $a/L_T = 3.3$, for a small-box simulation (green) and a large-box simulation (yellow), at the final simulation time in each case.
} 
\label{uncvgd_spectra}
\end{figure}

(ii) At the low end of the
collisionality scale in Figure \ref{evolution} (blue, red and green), it can be seen that there is no significant difference in the heat fluxes as the collisionality is varied,
leading to the plateau of hollow red squares in Figure \ref{qescan_uncvgd}\subref{qe_uncvgd} (and a similar plateau for the
other temperature gradient, blue squares). These runs are also excluded because of box-scale effects. In these
cases, somewhat counterintuitively, the small-box runs, which, however, extend to higher $k$'s, appear to be resolved and do not exhibit the plateau. Figure \ref{uncvgd_spectra} shows an example (the lowest-temperature-gradient and lowest-collisionality case), in which the perpendicular temperature is dominated by the box scale for the large-box run, which is excluded.

\begin{figure}[t]
\centering
\subfigure[]{
\includegraphics[scale=0.9]{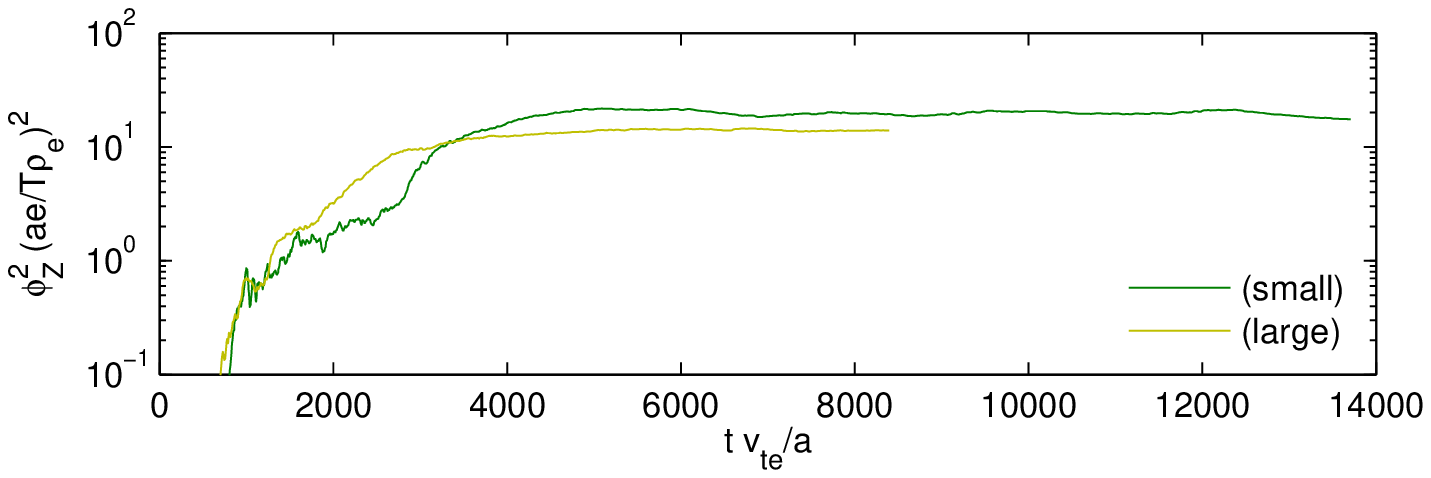} 
\label{zp_overlap}
}
\subfigure[]{
\includegraphics[scale=0.9]{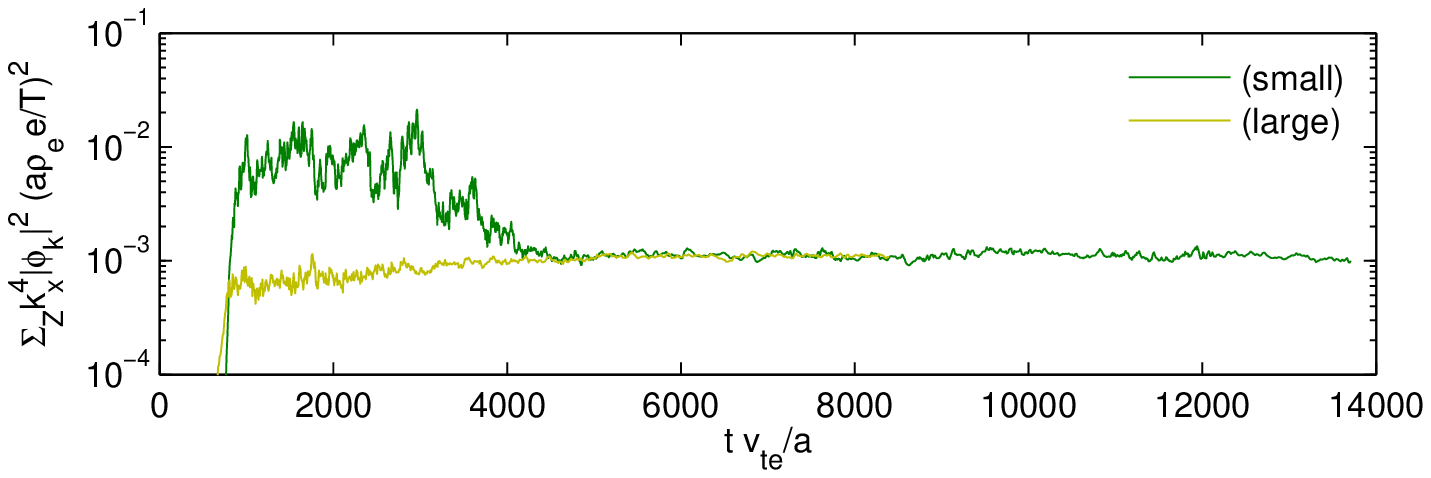}
\label{zs_overlap}
}
\subfigure[]{
\includegraphics[scale=0.9]{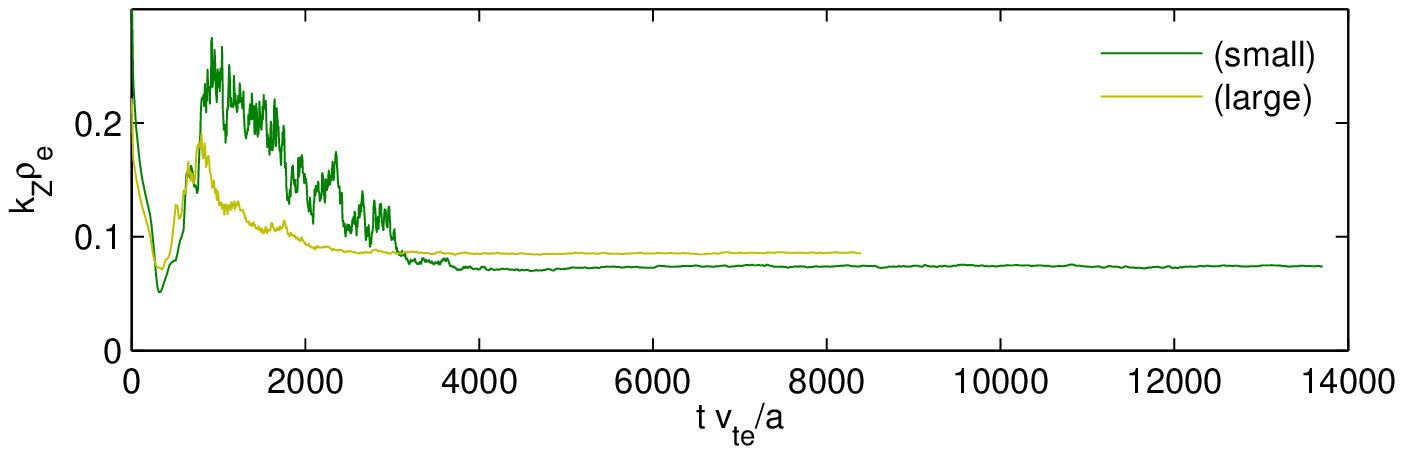}
\label{kz_overlap}
}
\caption{Evolution in time of \subref{zp_overlap} the square of the zonal electrostatic potential, $\phi_{\mathrm Z}^2 = \sum_{k_x}|\phi_{k_x,0}|^2$, \subref{zs_overlap} the square of the zonal shear, 
$\sum_{k_x} k_x^4|\phi_{k_x,0}|^2$, and \subref{kz_overlap} the rms zonal wavenumber $k_{\mathrm Z}$, for $\nu = 0.2 \,\nu_{\mathrm{nom}}$, $a/L_T = 3.3$ (green: small-box simulations; yellow: large-box simulations). This complements Figure \ref{overlap}.}
\label{overlap2}
\end{figure}

Eliminating both pathologies gives a fairly clean collisionality scaling when the remaining resolved runs are combined.
Given that the zonal damping, according to the scaling demonstrated in section \ref{zonal_damping_section}, is lowest for
the box-scale mode, it is not surprising that there are convergence problems at this end of the spectrum. Nonetheless, all the simulations that we consider resolved have saturated with the nonlinear state dominated by zonal modes with wavenumbers above the box-scale mode.

Thus, to summarise, at higher collisionalities, we need larger boxes but can sometimes get away with less spatial 
resolution, whereas at low collisionalities, we need
higher spatial resolution, but can sometimes get away with smaller boxes.

\begin{figure}[t]
\centering
\subfigure[]{
  \includegraphics[scale=0.65,clip]{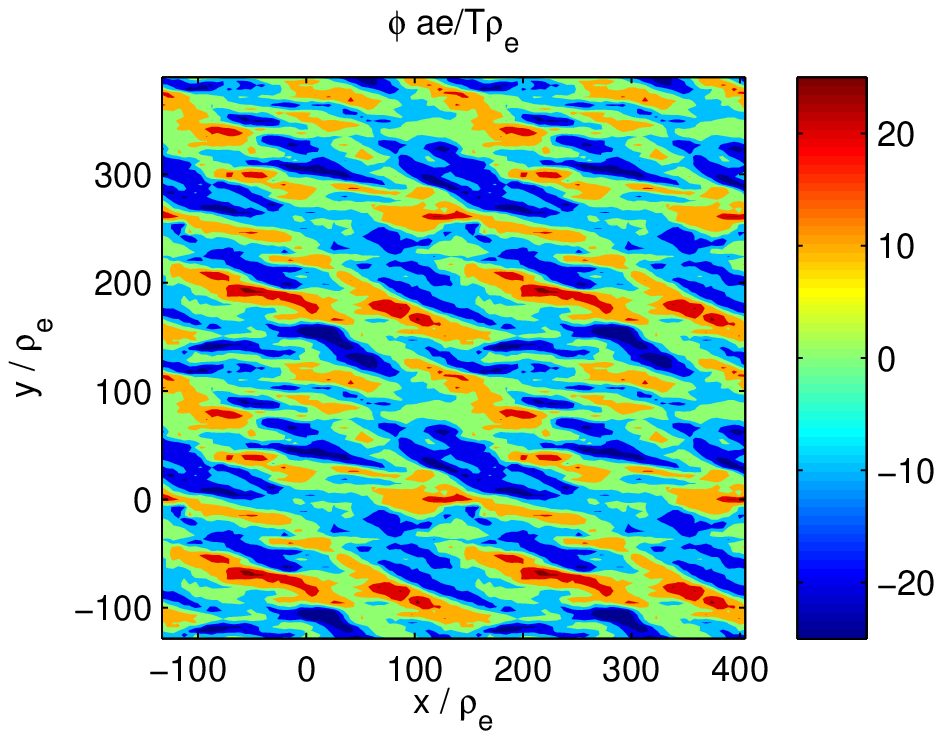} 
\label{quasisatslice2x2}
}
\subfigure[]{
  \includegraphics[scale=0.65,clip]{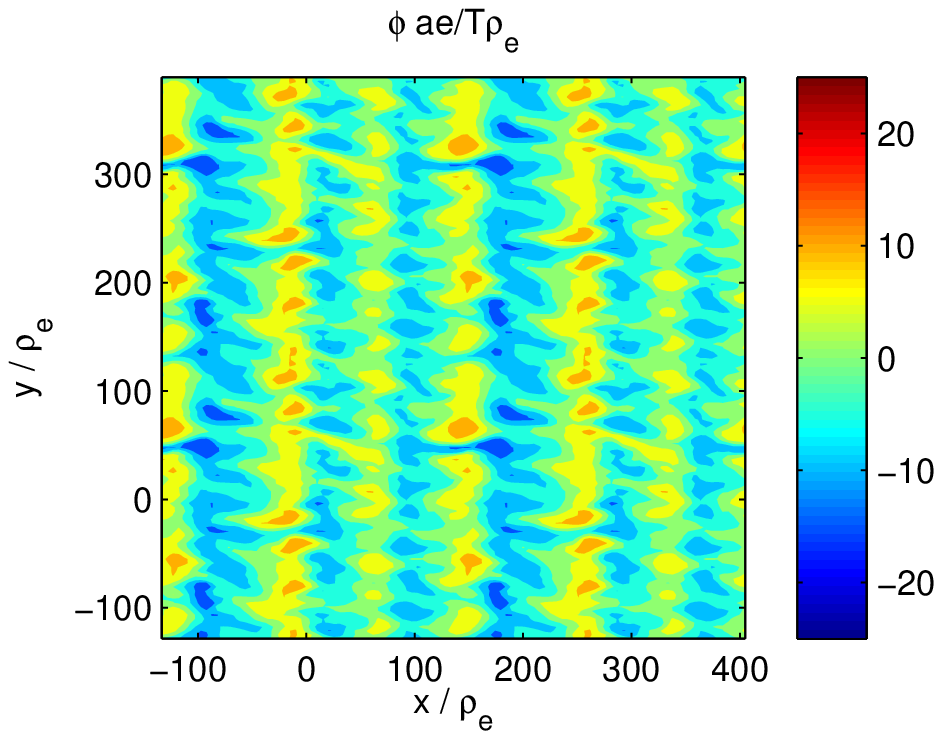} 
\label{satslice2x2}
}
\caption{Electrostatic potential $\phi$ at the outboard midplane, for $\nu = 0.2 \,\nu_{\mathrm{nom}}$, $a/L_T = 3.3$:  
\subref{quasisatslice2x2} quasi-saturated state at $t =  1204.2  \,a/v_{te}$, \subref{satslice2x2} long-time saturated state at $t =  7841.5  \,a/v_{te}$, for small-box simulations showing $2\times 2$ copies of each periodic domain. The large-box version of this simulation is shown in Figure \ref{contours}.}
\label{contours2}
\end{figure}

For completeness, we include as Figure \ref{overlap2} the time traces of various zonal quantities for the 
simulations used
in Figure \ref{overlap}. When these two figures are combined, the complete set analogous to Figure \ref{evolution} is obtained. Note that these demonstrate some lack of convergence in the ``quasi-saturated'' state, but the long-time state is resolved. 
Spatial cross-sections of the electrostatic potential for these two simulations are shown in Figure \ref{contours} (large-box) and Figure \ref{contours2} (small-box). The morphology of the quasi-saturated state is
qualitatively distinguishable between the two simulations. This difference is perhaps associated with the clear difference in the zonal shear shown in in Figure \ref{overlap2}\subref{zs_overlap}. 
The small-box simulation has higher spatial resolution, i.e., a higher maximum $k$, and the higher power of $k_x$ in the zonal shear (compared to the zonal velocity) emphasises the high end of the spectrum. The long-time saturated states, on the other hand, appear qualitatively the same. Note that in the large-box simulation in Figure \ref{satslice}, there are seven periods of the dominant zonal wavelength across the box, whereas in the doubled small box in Figure \ref{contours2}\subref{satslice2x2} there can only be an even number, which is six (two times the three across each smaller periodic box). This level of discrepancy between the zonal wavenumbers must be expected given the finite domain size.

\subsection{Background flow shear}
\label{shear}

Various authors have reported runaway growth in ETG turbulence simulations. 
To aid saturation, Guttenfelder and Candy~\cite{GuttenfelderCandy}, following earlier work \cite{CandyPPCF2007},
added background flow shear to their simulations. 
Roach et al.~\cite{Roach09}, extending the simulations of Joiner et al.~\cite{Joiner}, 
observed runaway growth that was healed by adding either flow shear or sufficient collisions 
(attributed by them to the stabilisation of weak trapped-electron modes close to the box scale 
at $k_y\rho_e\approx 0.02$). 

In our simulations, we have included an experimentally realistic level of flow shear, 
$\gamma_E= -0.003\,v_{te}/a$, and we do not find runaway behaviour. 
Note that in all of the cases reported above, the normalised perturbed zonal shear (Figure \ref{evolution}\subref{zs_evolution} and Figure \ref{overlap2}\subref{zs_overlap}) is an order of magnitude larger than the background flow shear. In the ``simplified'' simulations of section \ref{zeffscansection}, zero flow shear was used, and it is reasonable to suppose that the fixed electron-electron collisions helped regularise these simulations over a wider range of collisionalities than was accessible when varying electron-ion
and electron-electron collisions together. 
The simplified simulations support our view that the scaling of the heat flux with the collision rate that we have found does not depend directly on flow shear. 
Of course, physically, in a real tokamak, flow shear is still needed to suppress the ion-scale turbulence, which we did not model.

\section{Long-time linear damping of the electron zonal modes}
\label{Felix}

In this Appendix, we derive the scaling (\ref{gammaZ}) of the zonal damping rate. 

\subsection{Zonal evolution equation}

Consider the linearised form of the zonal gyrokinetic equation (\ref{GKE_Z}) (i.e., 
ignore its right-hand side). It is convenient to Fourier transform this equation 
in the field-perpendicular coordinate $x$ (there is no dependence on $y$  
by definition of the zonal modes), while leaving the field-parallel derivatives 
in position space (the zonal modes here are just ones with $k_y=0$, but are allowed 
to have parallel variation). This gives us
\begin{equation}
\frac{\partial}{\partial t}\left[h + J_0(k_\perp\rho)\frac{e\phi}{T} F\right] 
+ v_\parallel\mathbf{b}\cdot\nabla h + ik_x v_{Bx} h = 
\langle C[h e^{-ik_x\rho_x}]e^{ik_x\rho_x}\rangle,
\label{hzonal}
\end{equation}
where $J_0(k_\perp\rho)$ with $\rho = v_\perp/\Omega_e$ is the Fourier-space form of the 
gyroaveraging operator, $k_\perp = k_x|\nabla x|$ (here the curvilinear coordinate 
$x$ is defined by (\ref{defx}) and so $\nabla x = (q/B_0 r)\nabla\psi$), 
$\rho_x = (\nabla x)\cdot(\mathbf{b}\times\mathbf{v}_\perp)/\Omega_e$, the angle brackets 
$\langle\dots\rangle$ denote the averaging of the linearised collision operator over the gyroangle \cite{SCHEKOCHIHIN_APJS2009,Abel08}, and  
\begin{equation}
v_{Bx} = \mathbf{v}_{B}\cdot\nabla x = \frac{q}{B_0 r} \mathbf{v}_{B}\cdot\nabla \psi,
\label{vbx_def}
\end{equation}
where $\mathbf{v}_{B}$ is given by (\ref{vB}). 
The quasineutrality equation (\ref{QN2}) becomes 
\begin{equation}
\frac{e\phi}{T}\left(1 + \frac{1}{\tau}\right) = - \frac{1}{n}\int d^3\mathbf{v} J_0(k_\perp\rho) h. 
\label{phizonal}
\end{equation}

\subsection{Long-wavelength ordering and time scales}

Assuming that the zonal modes have long wavelengths, $k_x\rho_e\ll1$,  
we seek the solution to equations (\ref{hzonal}) and (\ref{phizonal}) in the form 
of an asymptotic expansion 
\begin{equation}
h = h^{(0)} + h^{(1)} + h^{(2)} + \dots,
\end{equation}
where $h^{(n)}\sim(k_x\rho_e)^n h^{(0)}$. 
In this ordering, $J_0(k_\perp\rho)\approx 1$ to lowest order and 
clearly, from (\ref{phizonal}), to lowest order, $e\phi/T \sim h^{(0)}/F$.  
We introduce a formal ordering of time scales in (\ref{hzonal}): 
\begin{equation}
\fl
v_\parallel\mathbf{b}\cdot\nabla \sim \frac{v_{te}}{a} \sim \nu,\quad
k_x v_{Bx} \sim \frac{k_x}{\Omega_e}\frac{v_{te}^2}{a} = k_x\rho_e \frac{v_{te}}{a} 
\sim (k_x \rho_e) \nu, \quad
\frac{\partial}{\partial t} \sim \gamma_\mathrm{Z} \sim (k_x\rho_e)^2\nu.
\label{ordering}
\end{equation}
This ordering anticipates the kind of solution we are expecting. 
Generally speaking, any initial zonal perturbation will evolve on three successive 
time scales: first, it will be damped collisionlessly by the streaming and 
magnetic-drift terms on the time scale $\sim a/v_{te}$, leaving a finite 
residual perturbation \cite{RH}; then this residual, which is non-Maxwellian, 
will be damped by collisions on the time scale $\sim 1/\nu$ 
\cite{HR,Kim,Xiao,Xiao07}, leaving a residual perturbed Maxwellian 
(this will be our $h^{(0)}$); and finally, this perturbed Maxwellian  
will be damped diffusively at the rate $\gamma_\mathrm{Z}$, as we are 
about to show. 

Thus, the zonal perturbation whose evolution we are going to calculate 
will be close to a steady state, with its evolution essentially governed 
by the (neoclassical) collisional transport theory. 

\subsection{Zeroth order}
\label{order0}

With the ordering (\ref{ordering}), at zeroth order, (\ref{hzonal}) becomes 
\begin{equation}
v_\parallel \mathbf{b}\cdot\nabla h^{(0)} = C[h^{(0)}]. 
\label{zerothorder}
\end{equation}
If we multiply this equation by $h^{(0)}/F$ and flux-surface average, 
the left-hand side vanishes (see, e.g., \cite{Abel13}, \S 6.1)
and we get 
\begin{equation}
-\overline{\int d^3\mathbf{v} \frac{h^{(0)} C_{ee}[h^{0}]}{F}} 
+ \overline{\int d^3\mathbf{v} \frac{\nu_{ei}v_{te}^3}{v^3 F}\frac{1-\xi^2}{2}
\left(\frac{\partial h^{(0)}}{\partial\xi}\right)^2} = 0,  
\label{vanishing}
\end{equation}
where we have denoted the flux-surface average by an overbar, used the collision 
operator (\ref{model_e}) taken to lowest order in $k_\perp\rho_e$ and 
integrated by parts in the term involving electron-ion collisions  
(cf.\ \cite{SCHEKOCHIHIN_APJS2009}, \S 4.2). Both terms in (\ref{vanishing}) 
are non-negative definite and so must vanish individually. 
The vanishing of the first of these implies that $h^{(0)}$ must be 
a perturbed Maxwellian, the vanishing of the second that this perturbed Maxwellian 
has no mean parallel flow. Therefore, 
\begin{equation}
h^{(0)} = \left[-\frac{e\phi}{T}\left(1 + \frac{1}{\tau}\right) 
+ \left(\frac{v^2}{v_{te}^2}-\frac{3}{2}\right)\frac{\delta T}{T}\right] F,
\label{h0}
\end{equation} 
where the density perturbation associated with $h^{(0)}$ has been fixed 
by the requirement that, to lowest order, $h^{(0)}$ and $\phi$ must satisfy (\ref{phizonal}). 
Note that we omit the superscript $(0)$ on $\phi$ and $\delta T$ because we will not need to calculate 
these fields explicitly to any higher order. 

Finally, substituting (\ref{h0}) back into (\ref{zerothorder}) and using the fact 
that, $h^{(0)}$ being a flow-less perturbed Maxwellian, $C[h^{(0)}]=0$, we find that both 
$\phi$ and $\delta T$ must be flux functions to zeroth order: 
\begin{equation}
\mathbf{b}\cdot \nabla\frac{e\phi}{T} = \mathbf{b}\cdot \nabla \frac{\delta T}{T} = 0.
\end{equation} 

The zeroth-order zonal solution (\ref{h0}) is precisely the promised 
quasi-steady state to which any initial zonal perturbations will relax after initial collisionless 
and collisional transients. 

\subsection{First order}
\label{order1}

At first order in $k_x\rho_e$, equation (\ref{hzonal}) is
\begin{equation}
v_\parallel\mathbf{b}\cdot\nabla h^{(1)} + ik_x v_{Bx} h^{(0)} = C[h^{(1)}].
\label{firstorder}
\end{equation}
Note that any gyroaveraging or FLR effects in the collision operator will only appear 
in the second order. 

In dealing with (\ref{firstorder}), it is convenient to take advantage of the fact that in 
tokamaks~\cite{HintonWong},
\begin{equation}
\mathbf{v}_B\cdot\nabla\psi = - v_\parallel\mathbf{b}\cdot\nabla\left(\frac{v_\parallel I}{\Omega_e}\right),
\label{drifttrick}
\end{equation}
where $I(\psi) = B_T R$ is a flux function, $B_T$ is the toroidal magnetic field and $R$ the major 
radius coordinate. Note that $\nabla$ on the right-hand side of (\ref{drifttrick}) is taken at constant $\mathcal{E}$ and $\mu$, which, owing to the variation of $B$, is not at constant $v_\parallel$. Using (\ref{vbx_def}) and recalling that the safety factor $q(\psi)$ and the 
radial flux-surface label $r(\psi)$ are both flux functions and the reference magnetic 
field $B_0$ is a constant, we have
\begin{equation}
v_{Bx} = - v_\parallel\mathbf{b}\cdot\nabla\left(\frac{v_\parallel}{\Omega_e}\frac{q I}{B_0 r}\right) 
= - v_\parallel\mathbf{b}\cdot\nabla\left(\rho_{pe} \frac{v_\parallel}{v_{te}}\right),
\label{vbx}
\end{equation}
where $\rho_{pe} = (qI/B_0 r)\rho_e$ is, by definition, the ``poloidal gyroradius'',
so called because $qI/B_0 r \sim B/B_p$, 
where $B_p$ is the poloidal magnetic field.   
Note that $\Omega_e$ and, therefore, $\rho_e = v_{te}/\Omega_e$, is calculated using 
the total, space-dependent field $B$, and so $\rho_{pe}$ is {\em not} a flux function.  

Since $h^{(0)}$ is a flux function ($\mathbf{b}\cdot\nabla h^{(0)}=0$), 
we may then rewrite (\ref{firstorder}) as follows
\begin{equation}
v_\parallel\mathbf{b}\cdot\nabla\left( h^{(1)} - ik_x\rho_{pe}\frac{v_\parallel}{v_{te}} h^{(0)}\right) 
= C[h^{(1)}].
\label{h1}
\end{equation}
The solution of this equation is a standard problem in neoclassical theory 
\cite{HintonWong,Helander}, but we shall not require its explicit form 
in order to work out the scalings that we seek. Note from (\ref{h1}) that 
the first-order zonal perturbation has a parallel electron flow 
(and, therefore, current) of order 
\begin{equation}
\frac{u_\parallel}{v_{te}} \sim k_x\rho_{pe} \frac{\delta p}{p}, 
\end{equation}
where $\delta p/p$ is the zonal pressure perturbation associated with $h^{(0)}$. 
The perpendicular $\mathbf{E}\times\mathbf{B}$ zonal flow is, of course, just 
\begin{equation}
\frac{v_{Ey}}{v_{te}} \sim k_x\rho_e \frac{e\phi}{T}.
\end{equation} 
It is the resistive damping of zonal flows (currents) that will lead to the damping 
of the zonal modes. 

\subsection{Second order}

In order to calculate this damping, we need evolution equations for the zeroth-order 
fields $\phi$ and $\delta T$. The time derivatives enter at the second order in (\ref{hzonal}). 
Using (\ref{h0}) and (\ref{vbx}), we have at this order: 
\begin{eqnarray}
\nonumber
& \frac{\partial}{\partial t}\left[-\frac{1}{\tau} \frac{e\phi}{T} 
+ \left(\frac{v^2}{v_{te}^2}-\frac{3}{2}\right)\frac{\delta T}{T}\right] F 
+ v_\parallel\mathbf{b}\cdot\nabla h^{(2)} - h^{(1)}  
v_\parallel\mathbf{b}\cdot\nabla\left(ik_x\rho_{pe}\frac{v_\parallel}{v_{te}}\right) \\
& = k_x^2 \langle C[h^{(0)}\rho_x]\rho_x\rangle + 
\left\langle C\left[h^{(2)} - \frac{1}{2}k_x^2\rho_x^2h^{(0)}\right] \right\rangle\vphantom{\biggl(},
\end{eqnarray}
where the leading-order FLR parts of the collision operator acting on $h^{(0)}$ 
have appeared. We now transit-average this equation, integrate 
the magnetic-drift term by parts and use (\ref{h1}) to express 
$v_\parallel\mathbf{b}\cdot\nabla h^{(1)}$. 
The result, with the transit-average denoted by overbar, is 
\begin{equation}
\fl
\frac{\partial}{\partial t}\left[-\frac{1}{\tau} \frac{e\phi}{T} 
+ \left(\frac{v^2}{v_{te}^2}-\frac{3}{2}\right)\frac{\delta T}{T}\right] F 
= \overline{-  i k_x \rho_{pe} \frac{v_\parallel}{v_{te}} C[h^{(1)}]  
+ k_x^2 \langle C[h^{(0)}\rho_x]\rho_x\rangle + \langle C[\dots] \rangle}.
\label{secondorder}
\end{equation} 

To separate the evolution of $\phi$ and $\delta T$, we take the density 
and energy moments of (\ref{secondorder}). 
Integrating it over velocities, and noting that,  
by conservation of particles, $\int d^3\mathbf{v}\langle C[\dots]\rangle = 0$, 
we get\footnote{The last term in (\ref{slow_phi}) is related to the spatial FLR diffusion term 
that we wrote out explicitly in the gyroaveraged collision operator (\ref{model_e}).}
\begin{equation}
-\frac{n}{\tau} \frac{\partial}{\partial t} \frac{e\phi}{T}  
= \overline{-i k_x \rho_{pe} \int d^3\mathbf{v} \frac{v_\parallel}{v_{te}} C_{ei}[h^{(1)}]
+ k_x^2\rho_e^2\int d^3\mathbf{v} \frac{C_{ei}[h^{(0)}\rho_x]\rho_x}{\rho_{e}^2}}.
\label{slow_phi}
\end{equation}
Only the electron-ion collision terms have survived because, to lowest order,  
conservation of momentum by the electron-electron collisions implies  
\begin{equation}
\int d^3\mathbf{v} v_\parallel C_{ee}[h^{(1)}] = 0, \quad
\int d^3\mathbf{v} C_{ee}[h^{(0)}\rho_x] \rho_x = 0.  
\end{equation}
In the same vein, multiplying (\ref{secondorder}) by $v^2/v_{te}^2$ and integrating over velocities, 
we find, using energy conservation by collisions, $\int d^3\mathbf{v} v^2\langle C[\dots]\rangle = 0$,  
\begin{equation}
\fl
\frac{3n}{2} \frac{\partial}{\partial t}\frac{\delta T}{T}  
= \overline{-i k_x \rho_{pe} \int d^3\mathbf{v} \frac{v_\parallel}{v_{te}}
\left(\frac{v^2}{v_{te}^2}-\frac{3}{2}\right) C[h^{(1)}]
+ k_x^2\rho_e^2\int d^3\mathbf{v} \frac{C[h^{(0)}\rho_x]\rho_x}{\rho_e^2}
\left(\frac{v^2}{v_{te}^2}-\frac{3}{2}\right)}.
\label{slow_dT}
\end{equation}  
The difference between this equation and (\ref{slow_phi}) is that the 
contributions from electron-electron collisions do not vanish 
(because same-species collisions can support non-zero heat fluxes). 

\subsection{Damping rate}

We do not need to solve the neoclassical equation (\ref{h1}) for $h^{(1)}$ explicitly 
to see that, this equation being linear, $h^{(1)}$ will be a linear combination of 
$e\phi/T$ and $\delta T/T$ with velocity-dependent coefficients, all of which 
are proportional to $k_x\rho_{pe}$; any homogeneous part of $h^{(1)}$ satisfies 
(\ref{zerothorder}) and so can be absorbed into $h^{(0)}$. 
Therefore, we may schematically represent (\ref{slow_phi}) and (\ref{slow_dT}) 
as the following matrix equation at this order:\footnote[1]{Note that in the absence of magnetic drifts, in a straight field, 
$h^{(1)}=0$ and we must replace $\rho_{pe}$ with $\rho_e$ everywhere, with the 
collisional evolution of the zonal modes now controlled by the FLR spatial-diffusion terms 
in the gyrokinetic collision operator.}
\begin{eqnarray}
\label{phi_schematic}
\frac{\partial}{\partial t} \frac{e\phi}{T} &= 
\nu_{ei} k_x^2\rho_{pe}^2\left( a_{11} \frac{e\phi}{T} + a_{12} \frac{\delta T}{T} \right)\\
\frac{\partial}{\partial t} \frac{\delta T}{T} &= 
\nu_{ei} k_x^2\rho_{pe}^2\left( a_{21} \frac{e\phi}{T} + a_{22} \frac{\delta T}{T} \right) 
+ \nu_{ee} k_x^2\rho_{pe}^2\left( b_{21} \frac{e\phi}{T} + b_{22} \frac{\delta T}{T} \right)
\label{dT_schematic}
\end{eqnarray}
where $a_{ij}$ and $b_{ij}$ are dimensionless coefficients that depend only 
on equilibrium parameters. Clearly, if $\nu_{ei}=0$, the matrix is defective (has a zero row corresponding to
the upper equation (\ref{phi_schematic})) and the damping rate of both the 
potential and temperature perturbations of the slowest damped eigenmode vanishes. Therefore, 
\begin{equation}
\gamma_\mathrm{Z} \propto \nu_{ei} k_x^2 \rho_{pe}^2. 
\end{equation}
The order-unity numerical prefator in the exact expression for $\gamma_\mathrm{Z}$ 
depends on the ratio $\nu_{ee}/\nu_{ei}$, which itself is order unity.  
Thus, we have shown that the long-time damping rate of the zonal modes always 
satisfies~(\ref{gammaZ}), a scaling that is indeed well reproduced in our 
numerical simulations (Figure \ref{zonal_damping}).

\subsection{Zonal damping by same-species collisions}
\label{intraspecies}

The result (\ref{gammaZ}) survives even if, artifically, the ratio $\nu_{ee}/\nu_{ei}$ 
is made large --- as was done in the ``simplified'' simulations of section \ref{zeffscansection}. 
In this case, (\ref{dT_schematic}) simply implies 
that 
\begin{equation}
\frac{\delta T}{T} \approx -\frac{b_{21}}{b_{22}}\frac{e\phi}{T}, 
\label{hardcouple}
\end{equation}
and, according to (\ref{phi_schematic}), 
the zonal perturbation is still damped at the 
rate $\gamma_\mathrm{Z} \sim \nu_{ei} k_x^2 \rho_{pe}^2$.

If we consider an even more artificial situation in which $\nu_{ei}$ is made to vanish 
completely, we must go to next order when calculating the density moment (\ref{slow_phi}) 
of the zonal kinetic equation (\ref{hzonal}) to allow higher-order finite-drift-orbit-width 
FLR terms to come in. These terms will be small by an extra 
factor of $(k_x\rho_{pe})^2$ and so technically we would have to reorder the time 
derivative two powers of $k_x\rho_e$ higher and then work through two 
extra orders in our expansion.\footnote{This was indeed confirmed 
independently by such a calculation \cite{Jack_note}.} It is clear though that the result of this 
amounts to replacing (\ref{phi_schematic}) and (\ref{dT_schematic}) with 
\begin{eqnarray}
\label{phi_schematic2}
\frac{\partial}{\partial t} \frac{e\phi}{T} &= 
\nu_{ee} k_x^4\rho_{pe}^4\left( b_{11} \frac{e\phi}{T} + b_{12} \frac{\delta T}{T} \right),\\
\frac{\partial}{\partial t} \frac{\delta T}{T} &= 
\nu_{ee} k_x^2\rho_{pe}^2\left( b_{21} \frac{e\phi}{T} + b_{22} \frac{\delta T}{T} \right),
\label{dT_schematic2}
\end{eqnarray}
using (\ref{dT_schematic2}) to infer again the coupling (\ref{hardcouple}) 
between the zonal potential and temperature perturbations, and finally arriving at 
a zonal damping rate 
\begin{equation}
\gamma_\mathrm{Z} \sim \nu_{ee} k_x^4\rho_{pe}^4. 
\end{equation}
This is the scaling that is indeed obtained numerically when $\nu_{ei}$ is switched off 
(black crosses in Figure~\ref{zonal_damping}).

These considerations might appear moot, as $\nu_{ei}=0$ is unphysical and achievable only in numerical experiments.  
They do, however, help us gain some insight into the way in which our theoretical aruments 
would be modified if we applied them to zonal dynamics in ITG rather than ETG turbulence, 
as suggested in section \ref{ITG}. 

\subsection{Case of ion zonal modes}
\label{ionzfs}

The only differences between the linearised gyrokinetic equation for the ion zonal modes 
and (\ref{hzonal}) are a minus sign in front of the $e\phi/T$ term, accounting for the ion charge,
and the absence of interspecies collisions: 
\begin{equation}
\frac{\partial}{\partial t}\left[h_i - J_0(k_\perp\rho)\frac{e\phi}{T_i} F\right] 
+ v_\parallel\mathbf{b}\cdot\nabla h_i + ik_x v_{Bx} h_i = 
\langle C_{ii}[h_i e^{-ik_x\rho_x}]e^{ik_x\rho_x}\rangle.
\label{hizonal}
\end{equation}
Another very important difference between the ion and electron cases 
is the Boltzmann-electron closure (\ref{ai2}), 
which removes the flux-surface-averaged part of the zonal potential from the electron-density 
response. Combining (\ref{ai2}) with the expression (\ref{dnn}) for the ion density 
perturbation in terms of $h_i$, we get the following version of the quasineutrality 
equation: 
\begin{equation}
\frac{e\phi}{T_i} + \frac{e(\phi - \overline{\phi})}{T_e}  
= \frac{1}{n}\int d^3\mathbf{v} J_0(k_\perp\rho) h_i. 
\label{QNizonal}
\end{equation}

That the difference between (\ref{QNizonal}) and (\ref{phizonal}) is significant 
becomes obvious if we take the density moment of (\ref{hizonal}), i.e., 
integrate it over velocities keeping $\mathbf{r}$ constant (equivalently, 
multiply by $J_0(k_\perp\rho)$ and integrate over $\mathbf{v}$), and then 
flux-surface average. To lowest order in $k_\perp\rho_i$, 
the term in the square brackets becomes, using (\ref{QNizonal}),  
\begin{equation}
\overline{\frac{1}{n}\int d^3\mathbf{v} J_0(k_\perp\rho) \left[h_i - J_0(k_\perp\rho)\frac{e\phi}{T_i} F\right]} 
\approx \frac{1}{2}k_\perp^2\rho_i^2 \frac{e\overline{\phi}}{T_i}.
\label{ddti}
\end{equation}
This has two extra powers of $k_x$ compared to the analogous term for electron zonal 
flows (see the left-hand side of (\ref{slow_phi})). 

By a calculation analogous to \ref{order0}, the zeroth-order solution $h_i^{(0)}$ 
of (\ref{hizonal}) in the long-time limit must again be a quasi-steady 
(i.e., slow-evolving), constant on flux surfaces, flow-less perturbed Maxwellian. 
The rest of the calculation 
proceeds similarly to the electron case with the exception that only momentum-conserving 
ion-ion collisions are present and so the velocity integral of the collision terms 
in the density equation will be smaller by an extra factor of $k_x^2\rho_{pi}^2$, 
as explained in \ref{intraspecies}. 
In view of (\ref{ddti}), the analog of (\ref{phi_schematic2}) for ion zonal modes will then be 
\begin{equation}
\frac{\partial}{\partial t}
\frac{1}{2}k_x^2\rho_{pi}^2 \frac{e\phi}{T_i} = \nu_{ii} k_x^4\rho_{pi}^4
\left(c_{11} \frac{e\phi}{T_i} + c_{12} \frac{\delta T_i}{T_i}\right),
\end{equation}
where we have absorbed all factors accounting for differences between 
$k_\perp$ and $k_x$ and between $\rho_i$ and $\rho_{pi}$ into the 
dimensionless coefficients $c_{11}$ and $c_{12}$.
The extra factors of $k_x^2\rho_{pi}^2$ on the left-hand side 
(due to the special role of flux-surface-averaged potential in the Boltzmann-electron closure) 
and on the right-hand side (due to momentum conservation by the ion-ion collision operator) 
cancel, leaving $e\phi/T_i$ to evolve on the diffusive time scale $\sim 1/\nu_{ii} k_x^2\rho_{pi}^2$. 
The evolution equation for $\delta T_i/T_i$ does not have 
these extra factors on either side and so feature the same time scale. Thus, the ion zonal 
modes, similarly to the electron ones (although for a different reason), 
will be damped at a rate 
\begin{equation} 
\gamma_{\mathrm{Z} i} \sim \nu_{ii} k_x^2 \rho_{pi}^2. 
\end{equation}

We remind the reader that the above calculation, like the whole of \ref{Felix}, is concerned solely with the linear physics of the zonal modes considered on their own. Its significance nonlinearly will depend upon the interactions between these modes and the nonzonal modes. In the main part of the paper, we report nonlinear results for ETG turbulence only, but we do discuss ITG turbulence further, in this wider nonlinear context, in the concluding section \ref{ITG}.

\section*{References}

\bibliography{long_etg_bibliography_nourls}

\end{document}